\documentclass[fleqn,usenatbib]{mnras}

\usepackage{newtxtext,newtxmath}
\usepackage[T1]{fontenc}

\DeclareRobustCommand{\VAN}[3]{#2}
\let\VANthebibliography\thebibliography
\def\thebibliography{\DeclareRobustCommand{\VAN}[3]{##3}\VANthebibliography}

\usepackage{graphicx}	% Including figure files
\usepackage{amsmath}	% Advanced maths commands
\usepackage{xcolor}     % colour text (remove before submission)
\usepackage{caption}
\usepackage{subcaption}
\usepackage{soul}

\definecolor{titlecol}{rgb}{0,0,1}
\definecolor{titlecol2}{rgb}{0,0.4,0}
\definecolor{titlecol3}{rgb}{0,0.5,0.8}
\definecolor{titlecol4}{rgb}{0.039,0.361,0.569} %teal++

\newcommand\galfit      {{\tt GALFIT}}

\newcommand\agndiscs    {\textsc{AgnDiscs}}
\newcommand\agndiscfin  {\textsc{AgnDiscFin}}
\newcommand\HST         {\emph{HST}}
\newcommand\inacdiscs   {\textsc{InacDiscs}}
\newcommand\inacdiscmatch   {\textsc{InacDiscMatch}}

\newcommand\fpsf        {$f_{\mathrm{psf}}$}
\newcommand\fpsfhst     {$f_{\mathrm{psf},\mathrm{HST}}$}
\newcommand\fpsfsdss    {$f_{\mathrm{psf},\mathrm{SDSS}}$}

\newcommand\mstar       {$M_{\ast}$}

\title[Merger-Free AGN and Bars]{The most luminous, merger-free AGN show only marginal correlation with bar presence}

\author[I. L. Garland et al.]{Izzy L. Garland,$^{1}$\thanks{E-mail: i.garland@lancaster.ac.uk}
Matthew J. Fahey,$^{1}$
Brooke D. Simmons,$^{1}$
Rebecca J. Smethurst,$^{2}$
Chris J. Lintott,$^{2}$ \newauthor
Jesse Shanahan,$^{3}$
Maddie S. Silcock,$^{1,4}$
Joshua Smith,$^{1}$
William C. Keel,$^{5}$
Alison Coil,$^{3}$
Tobias G\'eron,$^{2}$ \newauthor
Sandor Kruk,$^{6}$
Karen L. Masters,$^{7}$
David O'Ryan,$^{1}$
Matthew R. Thorne,$^{1}$
Klaas Wiersema$^{1}$
\\
$^{1}$Physics Department, Lancaster University, Lancaster, LA1 4YB, UK\\
$^{2}$Oxford Astrophysics, Department of Physics, University of Oxford, Denys Wilkinson Building, Keble Road, Oxford, OX1 3RH, UK\\
$^{3}$Center for Astrophysics and Space Sciences (CASS), Department of Physics, University of California, San Diego, CA 92093, USA\\
$^{4}$Centre for Astrophysics Research, University of Hertfordshire, College Lane, Hatfield AL10 9AB, UK\\
$^{5}$Department of Physics and Astronomy, University of Alabama, Box 870324, Tuscaloosa, AL 35404, USA\\
$^{6}$Max-Planck-Institut f\"ur extraterrestrische Physik (MPE), Giessenbachstrasse 1, 85748 Garching bei M\"unchen, Germany\\
$^{7}$Haverford College, 370 Lancaster Avenue, Haverford, PA 19041, USA
}

\date{Accepted XXX. Received YYY; in original form ZZZ}

\pubyear{2022}

\begin{document}
\label{firstpage}
\pagerange{\pageref{firstpage}--\pageref{lastpage}}
\maketitle

\begin{abstract}
The role of large-scale bars in the fuelling of active galactic nuclei (AGN) is still debated, even as evidence mounts that black hole growth in the absence of galaxy mergers cumulatively dominates and may substantially influence disc (i.e., merger-free) galaxy evolution.
We investigate whether large-scale galactic bars are a good candidate for merger-free AGN fuelling. Specifically, we combine slit spectroscopy and \emph{Hubble Space Telescope} imagery to characterise star formation rates (SFRs) and stellar masses of the unambiguously disc-dominated host galaxies of a sample of luminous, Type-1 AGN with $0.02<z<0.24$. 
After carefully correcting for AGN signal, we find no clear difference in SFR between AGN hosts and a stellar mass-matched sample of galaxies lacking an AGN ($0.013<z<0.19$), although this could be due to small sample size ($n_{\mathrm{AGN}}=34$). We correct for SFR and stellar mass to minimise selection biases, and compare the bar fraction in the two samples. We find that AGN are marginally ($\sim1.7\upsigma$) more likely to host a bar than inactive galaxies, with AGN hosts having a bar fraction, $f_{\mathrm{bar}}=0.59^{+0.08}_{-0.09}$ and inactive galaxies having a bar fraction, $f_{\mathrm{bar}}=0.44^{+0.08}_{-0.09}$. However, we find no further differences between SFR- and mass-matched AGN and inactive samples.
While bars \emph{could} potentially trigger AGN activity, they appear to have no further, unique effect on a galaxy's stellar mass or SFR.

\end{abstract}

\begin{keywords}
galaxies: disc -- galaxies: active -- galaxies: bar -- galaxies: star formation
\end{keywords}

%%%%%%%%%%%%%%%%%%%%%%%%%%%%%%%%%%%%%%%%%%%%%%%%%%

%%%%%%%%%%%%%%%%% BODY OF PAPER %%%%%%%%%%%%%%%%%%

% \tableofcontents

\section{Introduction}
There are still many fundamental open questions about the interplay between galaxies and the supermassive black holes (SMBHs) they host. For example, whilst major galaxy mergers were thought to dominate black hole-galaxy co-evolution in previous decades \citep[e.g.,][]{kormendy2013}, more recent results have made clear that merger-free (sometimes called `secular') processes are at least as important to the overall growth and evolution of black holes and galaxies as mergers{, with their black hole-galaxy properties showing the same correlations as those undergoing merger-driven growth \citep[e.g.,][]{simmons2017}, as described in more detail below}.

From the theoretical perspective, multiple cosmological simulations find that a dominant majority of black hole growth occurs as a result of merger-free processes \citep[at least 65 per cent, possibly more than 85 per cent, depending on the simulation;][]{martin2018,mcalpine2020}. Observational works have long been accumulating evidence for the merger-free black hole growth pathway \citep{greene2010, jiang2011b, cisternas2011, schawinski2011, kocevski2012, simmons2011, simmons2012, simmons2013, smethurst2021}, where often merger-free growth is isolated via the study of strongly disc-dominated galaxies \citep[which have not had a significant merger since $z \sim 2$;][]{martig2012}. 

Given the diversity of evidence for substantial merger-free black hole growth at a range of redshifts, there must be a significant mechanism of fuelling AGN in the absence of major mergers. In these secularly built, disc-dominated galaxies, gas must still be transported to the central regions in order for an AGN to be present. \citet{smethurst2019} calculate the necessary inflow rate {(i.e. the minimum gas mass that must be provided by any means to the central SMBH)} for their sample of AGN in disc-dominated galaxies, and show that bars {\citep{shlosman1989, shlosman1990, knapen1995, sakamoto1999, maciejewski2002, regan2004, lin2013}}, spiral arms \citep{maciejewski2004, davies2009, schnorrmuller2014}, and the smooth accretion of cold gas \citep{keres2005, sancisi2008} can each match the inflow rate required to sustain an AGN. These are all morphological features with a long lifespan \citep{miller1979, sparke1987, donner1994, donghia2013, hunt2018}, orders of magnitude longer than the $\sim 10^5\,\mathrm{yr}$ phases within the lifetime of an AGN \citep{schawinksi2015}, so if these features are able to periodically feed the SMBH \citep{schawinksi2015} over their lifetimes, then the mass of the SMBH can grow to the masses observed in the present. In other words, the secular, calm processes seen in disc-dominated galaxies are more than capable of fuelling AGN \citep{smethurst2019}.

Large-scale galactic bars, in particular, are a common feature in the local Universe, with \citet{masters2011} estimating that around $29.4\pm0.5$ per cent of disc galaxies at redshift $0.01<z<0.06$ host a large-scale, {strong} galactic bar when observed in optical wavelengths. {\citet{barazza2008} examine bar fraction in the $r$-band specifically, and report a bar fraction of 48 per cent to 52 per cent, however in infra-red imaging, a bar fraction as high as 70 per cent has been observed \citep{mulchaey1997, knapen2000, eskridge2000}.} Theoretical studies of AGN fuelling in disc galaxies show that bars are a viable method of transporting matter to a central SMBH {\citep{friedli1993, athanassoula1992, athanassoula2003, ann2005}}.

Despite bars being relatively common in disc galaxies and theoretically able to power a luminous AGN, observing such a connection has proven difficult. Many studies find no correlation between bars and AGN \citep{martini2003, oh2012, lee2012, cheung2015, goulding2017}, whereas studies such as \cite{knapen2000}, \cite{laine2002}, and \cite{laurikainen2004} show there is an increase in the number of AGN host galaxies containing bars of around 20 per cent. \citet{galloway2015} note that there is a higher probability of an AGN host galaxy possessing a {strong} bar than a galaxy without an AGN, but find no link between bars and the quantity or efficiency of AGN fuelling, indicating that whilst the presence of a {strong} bar may trigger the "turn on" of the AGN, the bar then drives accretion in a way that is indistinguishable from the secular processes that would be fuelling the AGN in the bar's absence.

%Several factors likely contribute to the difficulty of observing a connection between AGN and bars. AGN are more likely to reside in galaxies with a higher stellar mass, \mstar\ \citep{kauffmann2003, aird2012}, as are bars \citep{skibba2012}. Bars are also more likely to reside in redder galaxies (i.e., less star-forming) \citep{masters2011, masters2012, skibba2012, oh2012, cheung2013, geron2021}. Controlling for these confounding variables in order to understand how bars, star formation, and black hole growth may affect each other requires large samples and careful measurements.
Several factors likely contribute to the difficulty of observing a connection between AGN and bars. AGN are more likely to reside in galaxies with a higher stellar mass, \mstar\ \citep{kauffmann2003b, aird2012}, {and a correlation between bars and stellar mass has been observed, although the nature of this correlation is debated, potentially with bars being more likely to reside in galaxies with a higher stellar mass \citep[e.g.][]{skibba2012}, although a study conducted in \citet{erwin2018} highlights that this may be a selection effect, and shows that bar presence may peak at $\log(M_{\ast}/M_{\odot}) = 9.7$ and decrease either side of this value.} Bars are also more likely to reside in redder galaxies (i.e., less star-forming){ \citep{masters2011, masters2012, skibba2012, oh2012, cheung2013, kruk2018, geron2021}, but in some cases, enhancement of star formation rate (SFR) in a circumnuclear ring at the centre has been observed \citep{hawarden1986}}. Controlling for these confounding variables in order to understand how bars, star formation, and black hole growth may affect each other requires large samples and careful measurements.%but in some cases enhancement in centre (1986)

There is another crucial caveat in determining any link between bars and AGN which causes significant problems: both features have drastically different typical lifetimes. SMBH tend to only be in the AGN phase for around $10^5\,\textrm{yr}$ \citep{schawinksi2015}, whereas bars are long-lived features {\citep{sellwood2014}, with simulations showing bars that form at $z\sim1.3$ can maintain their strength down to $z\sim0$ \citep{kraljic2012}}. This corresponds to a lookback time of $8.9 \mathrm{Gyr}$, meaning that bars can live at least 100,000 times as long as an AGN phase. This means that when a barred galaxy is observed, we may not observe AGN activity because the AGN has since faded. Since bars tend to facilitate the development of pseudobulges over time via the buckling of stellar orbits (see \citealt{combes2009} for a review), observing galaxies with no or very small bulges may aid in mitigating this issue, as then any bars observed would be younger, and have less chance of outliving an AGN at the time of observation.

There is also very little consensus on the link between AGN and SFR \citep[e.g.][]{mulcahey2022}. Additionally, it is a challenge to measure SFRs in galaxies hosting luminous AGN. Star formation and AGN appear to share a common fuel source \citep{silverman2009}; thus if there is more of this fuel source, we would expect to see an increase in AGN and in SFR appearing together. This has been observed \citep[e.g.][]{mullaney2012, aird2019}. However, AGN feedback has also been shown to be capable of affecting the star formation in the host galaxy. For example, positive feedback can occur when an outflow compresses the molecular clouds or the interstellar medium in its path, thus increasing SFR \citep{ishibashi2012, schaye2015}. Negative feedback can quench star formation via heating the molecular gas and interstellar medium \citep[e.g.][]{ciotti2010}. See \citet{fabian2012} for a review on AGN feedback and star formation.

In this work, we examine AGN in unambiguously disc-dominated (`bulgeless') galaxies in order to isolate SMBH growth in the merger-free regime. {As mentioned above, these disc-dominated galaxies indicate a merger-free history since at least $z\sim2$, due to mergers resulting in a central bulge \citep{martig2012}.} Previous studies have shown that these AGN exist at a range of black hole masses and luminosities, \citep{satyapal2009, simmons2013, bizzocchi2014, satyapal2016}. \citet*{simmons2017} compiled a sample of relatively nearby ($z < 0.25$) unobscured, luminous AGN residing in disc-dominated systems. Despite having long-term evolutionary histories free of significant mergers, these systems lie on SMBH--galaxy co-evolution relations which were originally observed in elliptical galaxies with a history of major mergers \citep{haring2004}. This unique sample of merger-free quasars is the parent sample for the data used in this work. There has not yet been a detailed study of bars and AGN in these systems in the same way that there has been in the general galaxy population.

We use spectra taken from the Shane Telescope at Lick Observatory to examine the SFRs in merger-free galaxies hosting luminous AGN. We also investigate whether, after controlling for parameters such as SFR and \mstar, a correlation can be observed between the presence of a bar and the presence of an AGN. We discuss data collection, comparison samples and fitting procedures in Section \ref{sec:samp_and_obs}, and we determine stellar properties of our sample in Section \ref{sec:stellar_props}. We discuss SFR in Section \ref{sec:sfr_only}, and then examine the bar fractions in Section \ref{sec:controlling_for_sfr_and_mass}, before concluding in Section \ref{sec:conclusions}.

Throughout this paper, the term `active galaxy' refers to a galaxy that hosts an AGN, and the term `inactive galaxy' refers to a galaxy that does not host an AGN. These two terms do not refer to the star formation in the galaxy. We use WMAP9 cosmology \citep{hinshaw2013}, where we assume a flat universe, $H_0 = 69.3\,\mathrm{km}\,\mathrm{s}^{-1}\,\mathrm{Mpc}^{-1}$ and $\Omega_m = 0.287$.

\section{Sample and Observations}\label{sec:samp_and_obs}

This study uses multiple samples and data sources. In the subsections below, we describe our main sample of AGN host galaxies, as well as our comparison sample of inactive disc galaxies. We further describe the data reduction, spectral fitting, and morphological fitting procedures used for each of these samples.

\subsection{AGN host Sample}\label{sec:ssl_sample}
In order to investigate SMBH growth in the merger-free regime, we require a sample of AGN hosted in disc-dominated galaxies with little--to--no bulge component. The sample used here was first compiled in \citet{simmons2017}, and we summarise the sample selection here.

The initial sample of AGN is selected using the W2R sample \citep{edelson2012}, which were identified via a multi-wavelength approach using the Wide-Field Infrared Survey Explorer \citep[WISE;][]{wright2010}, Two Micron All-Sky Survey \citep[2MASS;][]{skrutskie2006} and the ROSAT All-Sky Survey \citep[RASS;][]{voges1999}. This photometric, all-sky selection combines both infrared and X-Ray selection to identify 4,316 unobscured AGN. \citep{edelson2012}. \citet{simmons2017} use the Sloan Digital Sky Survey \citep[SDSS;][]{York2000} to select from the AGN sample a set of galaxies that are dominated by the presence of a disc. Using SDSS Data Release 8 \citep[DR8;][]{aihara2011}, there are 1,844 sources within 3 arcsec of a source in the W2R sample. A single expert classifier (BDS) used the SDSS colour images to perform a morphological selection, and found that there were 137 galaxies lacking visual evidence of a bulge component, but containing features commonly found in discs (spiral arms, bars etc.). Many of these galaxies have SDSS fibre spectra focused on the nuclei of each source. However, in order to reliably determine SFRs in these Type-1 AGN with very strong emission lines, we require off-nuclear spectra.

\begin{figure}
    \centering
    \includegraphics[width=\columnwidth]{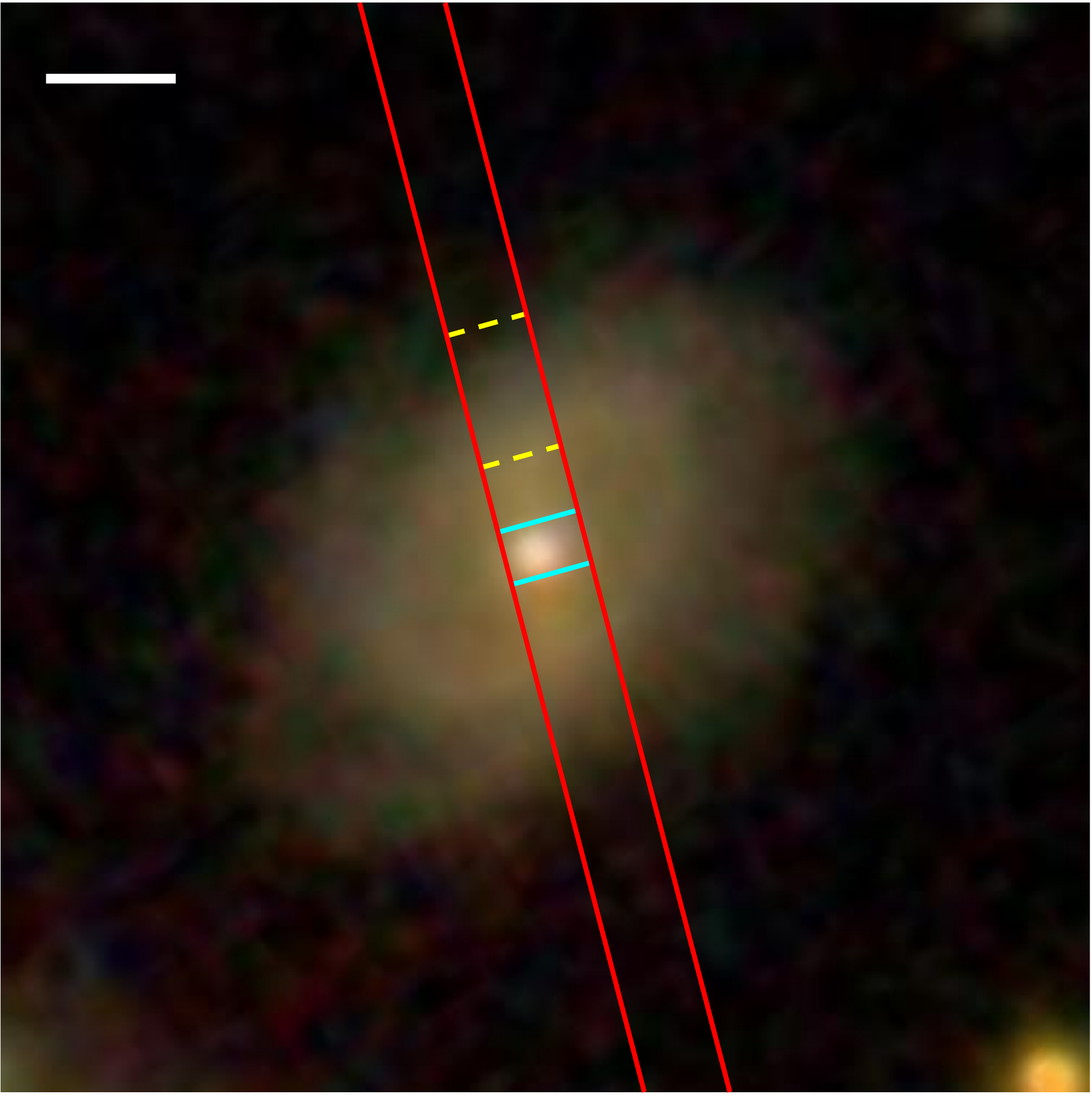}
    \caption{SDSS postage stamp of J081324.00+542236.9, overlain with the observed region, a slit of length 145 arcsec, shown as a red rectangle. The teal lines denote the 1D spectrum extracted from the central 5 arcsec of the slit, corresponding to the central spectrum shown in Figure \ref{fig:full_eg_spec}. The yellow lines denote the 1D spectrum extracted over the galaxy disc and is also shown in Figure \ref{fig:fpsf}. The scale bar shown in the top left corner corresponds to 10 arcsec.}
    \label{fig:sdss_obs_region}
\end{figure}

Longslit spectroscopic data was taken from the Kast Spectrograph on the Shane Telescope at Lick Observatory over 18 nights in the period 2016 October to 2018 November for {56} of these sources{, in order to work towards spectroscopic completion of the parent sample. Despite the 18 nights on sky, we were unable to obtain full spectroscopic completion of the sample, and 4 of these 137 sources have neither SDSS fibre spectra nor Lick longslit spectra. 21 of the sources have both longslit and fibre spectra}. Throughout this work, this sample of {56} sources shall be referred to as \agndiscs, and SDSS images of these sources are shown in Appendix \ref{app:sdss_imgs}.

\subsection{Inactive Sample}\label{sec:inactive_sample}
In order to investigate bar-driven fuelling of AGN, it is necessary to compare the AGN host sample to a resolution-matched {and morphology-matched} sample of galaxies which lack AGN activity signatures but are otherwise similar. This allows us to separate out any properties that may appear to be a result of bar presence, but are actually a result of AGN presence, as well as provide a baseline comparison for how a bar can affect a galaxy in the absence of an AGN. {Typically, when selecting comparison samples, stellar mass is also matched, and whilst we do perform this matching later on in Section \ref{sec:sfr_only}, we first want to see how the stellar mass (along with the star formation rate) varies between \agndiscs\ and the inactive galaxies.}

We used Galaxy Zoo 2 \citep[GZ2;][]{willett2013} to first identify a sample of disc-dominated galaxies. Volunteers are shown an image from SDSS, and asked via the question tree shown in \citet{willett2013} to classify the central galaxy in the image. The first question asked is `Is the galaxy smooth and rounded, with no sign of a disc?', and for this work, we require that the vote fraction for those who answered that the galaxy is featured be $p_{\mathrm{features\, or\, disc}} \geq 0.35$, following the suggestion in \citet{galloway2015} based on expert visual inspection. This leads the volunteers who answered `No' (i.e. the galaxy is featured) to the question `Could this be a disc viewed edge on?'. We require a sample of face-on discs so that we can identify a bar if one is present. In an edge-on disc, the bar is often hidden by the geometry of the galaxy. We require that the vote fraction of volunteers classing the disc as not-edge-on be $p_{\mathrm{not\, edge\, on}} \geq 0.6$, again following the suggestion in \citet{galloway2015}. This makes up our inactive disc sample.

To establish the lack of AGN, we use the fluxes from OSSY \citep{oh2011b} to divide the sample into AGN hosts, star-forming galaxies, composite sources, and LINERs. To build the inactive sample, we exclusively use sources that fall into the star-forming category. This is to ensure purity of the sample. We exclude any source where the emission lines [\ion{O}{iii}], [\ion{N}{ii}], H$\upalpha$, and H$\upalpha$ have a signal--to--noise ratio, $S/N < 3$. We use the guidance in \citet[][Equation 1]{kauffmann2003b}, where they show that a source is star-forming if

\begin{equation}
    \log\left([\ion{O}{iii}]\lambda6584 / \mathrm{H}\beta\right) < \frac{0.61}{\log\left([\ion{N}{ii}]\lambda6584 / \mathrm{H}\alpha\right) - 0.05} +1.3
\end{equation}

We impose a limit on the resolution rather than the redshift, since the bars are identified visually. We need to ensure the resolution distribution of active galaxies covers the same range as our sample of inactive galaxies. This is particularly important given that the inactive galaxies have their bar presence determined through SDSS images (via GZ2 volunteers), but only around half of the active galaxies use SDSS for bar identification - the rest use \HST\ images, which have a far better resolution and thus can push to higher redshift before the classification of bar presence is marred by significant doubt - see Section \ref{sec:bar_presence} for a more detailed description of identifying bars. For AGN hosts with \HST\ images, we determine what their equivalent redshift would be if they were observed solely with SDSS to obtain the same resolution in arcseconds per pixel. We use these equivalent redshifts to determine that the maximum redshift of our inactive sample should be $z > 0.187$. Ensuring this resolution matching is completed negates any issues that arise when identifying bars at different resolutions. After removing all inactive discs with $z > 0.187$, we are left with our comparison parent sample of 26,899 galaxies, which we refer to below as \inacdiscs.

\subsection{Data Reduction and Fitting}

\subsubsection{Lick Data Reduction}

We used the Image Reduction and Analysis Facility \citep[\textsc{Iraf};][]{tody1986, tody1993} to reduce the longslit \agndiscs\ spectra, and its packages designed specifically for longslit data reduction, \textsc{noao.twodspec.longslit}, and \textsc{noao.twodspec.apextract}. The Kast spectrograph has a red CCD and a blue CCD, and these were reduced separately. The instrument settings for all runs were: dichroic d57; blue grating 600/4310, red grating 600/7500. The slit width ranged from 2--3 arcsec, with a wider slit used for nights with particularly poor seeing. The overscan regions were subtracted, and the images were bias-subtracted and flat-fielded. There were a number of images, particularly in the red side of the detector, which were contaminated with cosmic rays, and for spectra taken in October 2016, stray alpha particles from a slightly radioactive instrument component that was later replaced. These artefacts were removed, and the images were calibrated for wavelength, then stacked according to the object and position angle. The background noise was subtracted from each combined image, and the images were extinction corrected. Standard stars, from which data was taken regularly throughout the night, were used to calibrate the flux at each wavelength. The standard stars used were: BD332642, BD284211, BD262606, Feige 34, Feige 110, G191B2B, G193-74, G24-9, GD248, HD157881, HD183143, HD19445, HD84937, HZ4. We use these standard stars to determine the point spread function (PSF) of the sources observed at that time. Since the standard stars are point sources, but have a Gaussian flux profile when observed, we can take the PSF to be the full--width--half--maximum (FWHM) of the flux of the star when plotted as a 2D spectrum.

\begin{figure*}
    \centering
    \includegraphics[width=\textwidth]{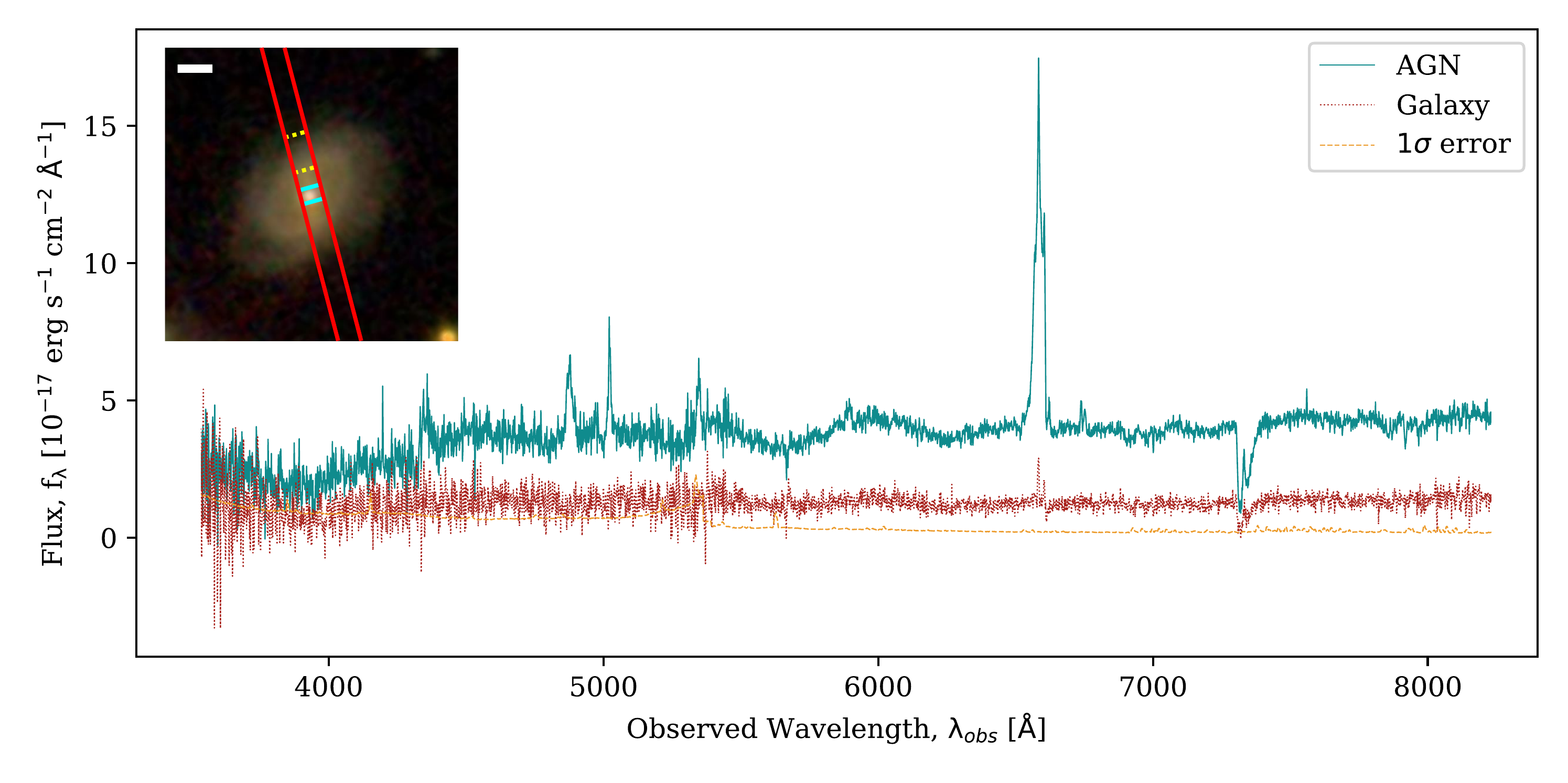}
    \caption{Full example spectra of J081324.00+542236.9 with AGN (solid teal line) and galaxy spectra (dotted red line) shown, and variance in the galaxy spectrum (dashed orange line). The thumbnail in the top left corner shows the galaxy from which these spectra were taken, and the red lines on the thumbnail represent the part of the image observed by the slit. The spectrum shown in red dashes is the spectrum taken over the galaxy, excluding a significant amount of the flux from the AGN. This corresponds to the section of the slit enclosed in neon yellow dashed lines The spectrum shown in solid blue is the spectrum taken over the central five pixels of the source, which is dominated by the flux from the AGN. This corresponds to the section of the slit encased in solid neon blue. The H$\upalpha$/[\ion{N}{ii}] is easily detected in both spectra, with an additional broad H$\upalpha$ component in the AGN spectrum. The [\ion{O}{iii}] and H$\upbeta$ emission lines are not apparent in the galaxy spectrum, but can be clearly seen in the AGN spectrum.}
    \label{fig:full_eg_spec}
\end{figure*}

Using longslit spectra means we can extract spectra at many points across the observed region, and we do this to obtain a spectrum of the central AGN in each source as well as an off-nuclear spectrum of the galaxy. The required 1D spectra were extracted; the 5 pixels around the central AGN to form the AGN spectrum, and the galaxy from either 2$\upsigma$ or 3$\upsigma$ of the PSF out to the edge of the disc to form the galaxy spectrum. Following reduction and extraction, the blue and red CCD outputs were merged to give two full spectra per position angle per object -- one of the galaxy and one of the AGN. Since the two sides of the detector each have a different spectral resolution, it is necessary to interpolate the region where the CCDs overlap. We aperture correct the AGN spectra to account for cases where the width of the slit is small compared to the PSF of the AGN. We assume that the central spectrum is dominated by AGN flux. This is due to our sample being selected so as to be the most luminous AGN. The slit and extraction regions are demonstrated in Figure \ref{fig:sdss_obs_region} for galaxy J081324.00+542236.9. We show the resultant spectra of J081324.00+542236.9 in Figure \ref{fig:full_eg_spec}, including an AGN spectrum, a galaxy spectrum and a variance spectrum.

\subsubsection{Spectral Fitting} \label{sec:spectral_fitting}
\begin{figure*}
    \centering
    \begin{subfigure}{0.47\textwidth}
        \centering
        \includegraphics[width=\columnwidth]{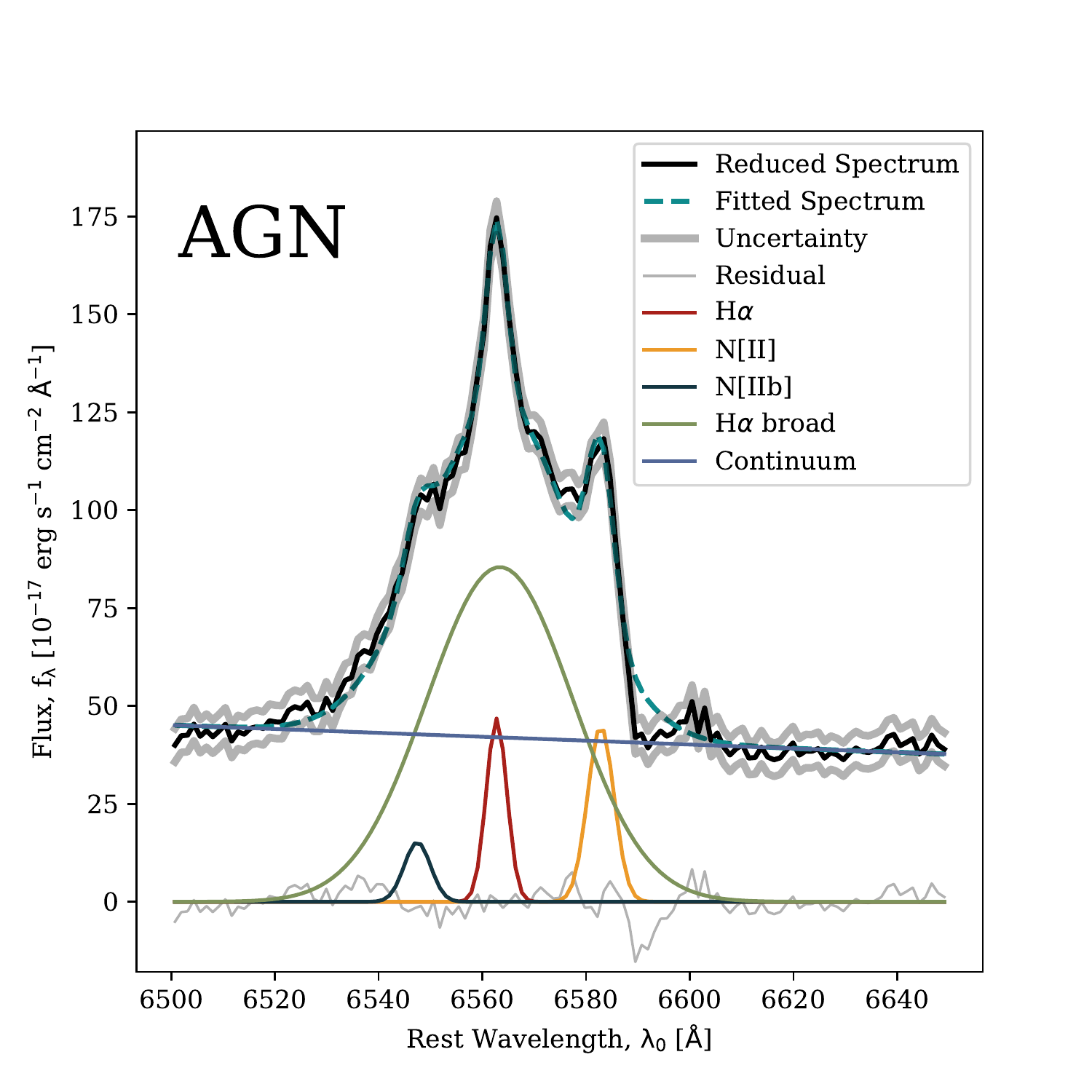}
        \caption{}
        \label{fig:eg_hanii_agn}
    \end{subfigure}
    \begin{subfigure}{0.47\textwidth}
        \centering
        \includegraphics[width=\columnwidth]{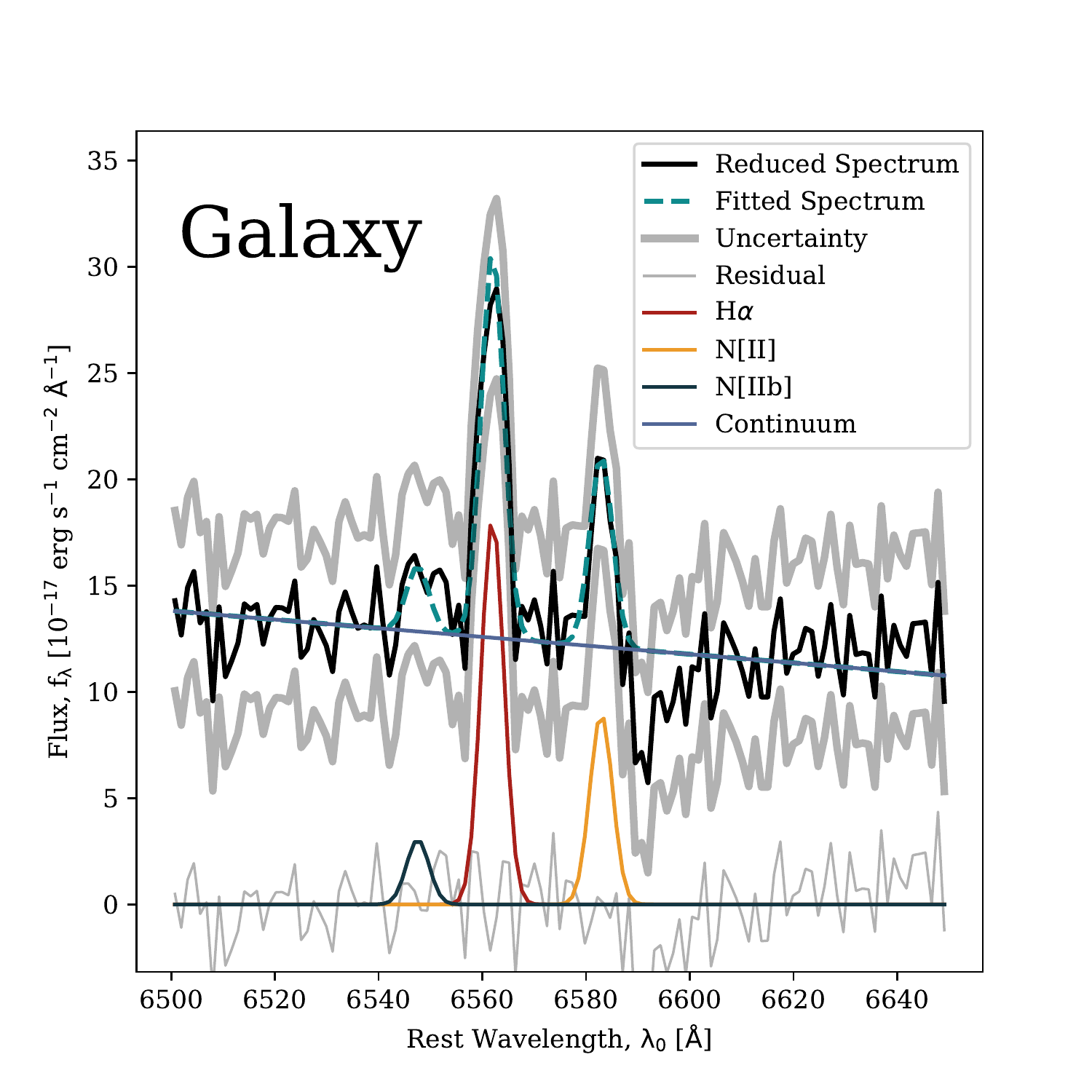}
        \caption{}
        \label{fig:eg_hanii_off_nuc}
    \end{subfigure}
    \caption{Fitted spectra, with panel \ref{fig:eg_hanii_agn} showing the spectrum across the centre of the source including the AGN, and panel \ref{fig:eg_hanii_off_nuc} showing the spectrum across the galaxy. The reduced spectrum is shown in black solid lines, and the fitted spectrum is shown in dashed turquoise, with the uncertainty in grey thick lines and the residual in grey thin lines. The components making up the fit are also shown, with the continuum in blue H$\upalpha$ in red, [\ion{N}{II}] in yellow, [\ion{N}{II}b] in dark blue, and broad H$\upalpha$ in green (only present in the AGN spectrum). The AGN spectrum primarily differs from the galaxy spectrum by the addition of this broad H$\upalpha$ component.}
    \label{fig:eg_hanii}
\end{figure*}

To fit the spectra, we used \textsc{Scipy} \citep{virtanen2020}, to fit a Gaussian function to each emission line along with a linear fit for the continuum emission near the line. The focus was on obtaining robust H$\upalpha$ and [\ion{O}{iii}] fits. For regions such as the H$\upalpha$/[\ion{N}{ii}] complex, several Gaussian functions were used to disentangle overlapping emission lines, as shown in Figure \ref{fig:eg_hanii}.

Where the signal--to--noise ratio was too low and we could not obtain accurate H$\upalpha$ fits of the sources, we determined the upper limit of H$\upalpha$ flux by assuming all the flux in the region where a detectable H$\upalpha$ emission line would have been is due to H$\upalpha$, and integrating the spectrum in this range to give a conservative upper limit.

The spectra taken over the centre of the system differ greatly to those taken of the galaxy. This is due to the presence of the AGN, which can add considerable flux and cause broadening. Thus, for all the AGN spectra, we require an extra Gaussian component for H$\upalpha$ with a higher velocity dispersion than the corresponding narrow component. This broad H$\upalpha$ component was also present in some of the off-nuclear spectra, and so was included in the fitting process since the AGN contaminant requires fitting before its successful removal. The differences in the galaxy and AGN spectra can be seen in Figure \ref{fig:eg_hanii}, with the AGN spectrum shown in Figure \ref{fig:eg_hanii_agn} and the galaxy spectrum shown in Figure \ref{fig:eg_hanii_off_nuc}. 

Redshifts were calculated using spectral emission lines. We used the [\ion{O}{iii}] 5007$\textrm{\AA}$ emission line as the reference wavelength where possible, however if for reasons such as low signal--to--noise the [\ion{O}{iii}] 5007$\textrm{\AA}$ observed wavelength was unreliable, we used the H$\upalpha$ 6563$\textrm{\AA}$ emission line.

After fitting the galaxy spectra, the AGN contaminant was subtracted. We observe that the Shane/Kast PSF is Gaussian by examination of standard star spectra. Thus where we extracted the galaxy spectrum from 2$\upsigma$ away from the AGN to the edge of the disc, we subtract 2.5 per cent of the AGN emission from the galaxy emission (since it is only one side of the PSF in the slit). Where instead we start at 3$\upsigma$, we subtract 0.015 per cent of the AGN emission. This gives us a final AGN host galaxy sample of {56} galaxies, {22} of which have upper limits constraining their H$\upalpha$ fluxes. This sample, which we refer to below as \agndiscs, has median redshift $z_{\rm med} = 0.0857$.

\subsubsection{\HST\ Data Reduction and Photometric Fitting}\label{sec:HST_data}

A subset of the AGN host galaxies selected via the method described above and analysed here were also observed with the \emph{Hubble Space Telescope} (\HST) Advanced Camera for Surveys (ACS) as part of a snapshot programme (HST-GO-14606, PI: B. Simmons). {Given that it was a snapshot programme, we prioritised those galaxies whose morphology was less clear in SDSS photometry, in order that confident morphologies could be obtained for all of \agndiscs\, as well as accurate decomposition of AGN and host galaxy, which we cover in more detail below.} Each of the 43 systems in \agndiscs\ with \HST\ imaging was observed in a single broadband optical filter, chosen to minimise the contribution of bright AGN emission depending on the redshift of the source (i.e., to avoid either [\ion{O}{iii}] /H$\upbeta$ or H$\upalpha$; typically this choice resulted in selecting the $F814W$ filter). 

Each source was observed with 2 short exposures to ensure an unsaturated nuclear PSF, and 2 long exposures to reach an acceptable depth in the extended galaxy. A typical exposure time on source was approximately 40 minutes, with ACS/WFC subarrays chosen to minimise readout time whilst still imaging substantial sky background. The data was reduced using the standard reduction pipeline\footnote{At the time of data reduction, some manual steps were required as a result of using subarrays, but these configurations have since been incorporated into the standard imaging reduction pipeline.}, including CCD charge diffusion correction and cosmic ray removal using \texttt{LACosmic} \citep{vanDokkum2001}. The long exposures were combined into a final science exposure. For the purposes of photometric fitting (described below), image fluxes of the reduced images are in counts.

\begin{figure}
    \centering
    \includegraphics[width=\columnwidth]{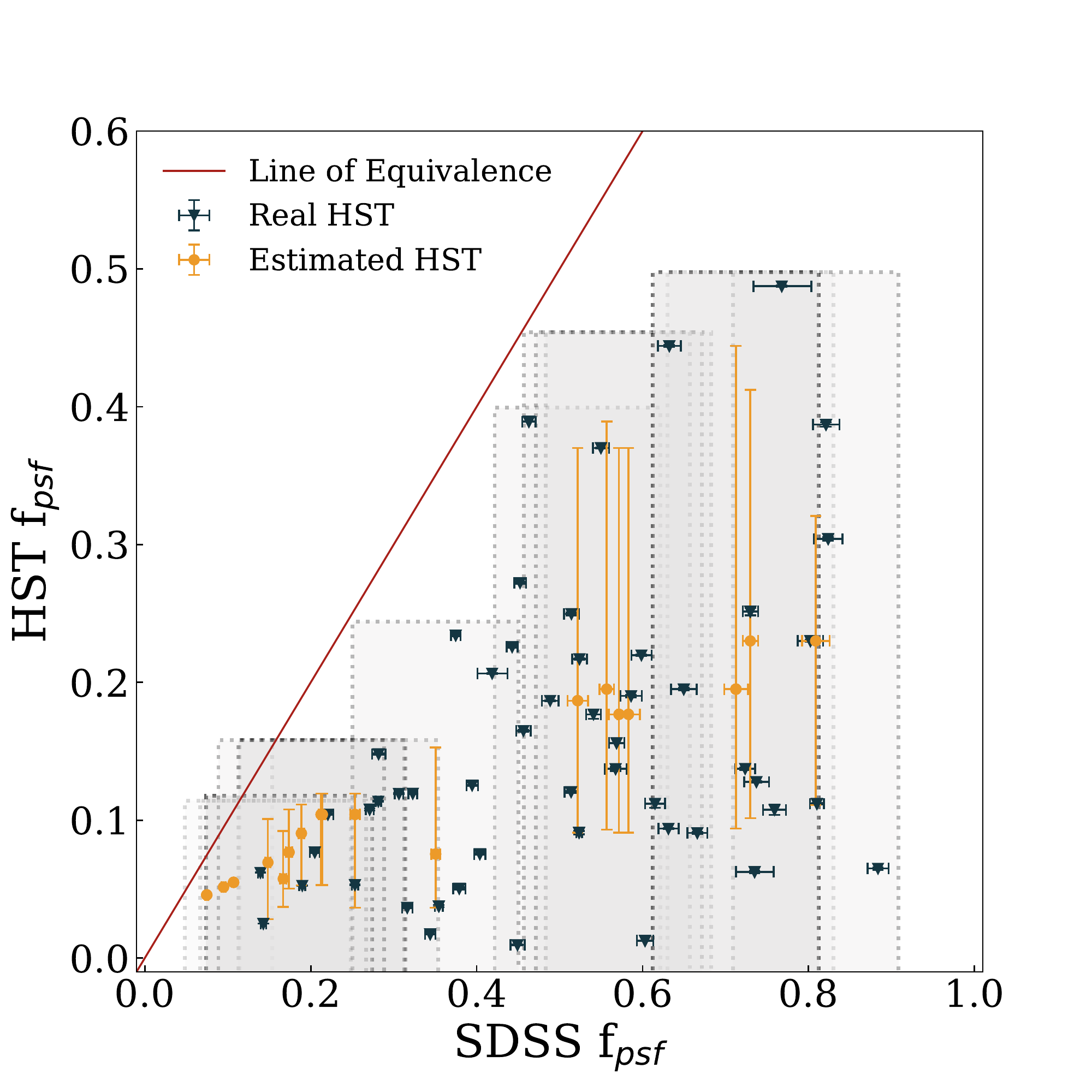}
    \caption{The fraction of the flux contained within the PSF for both \HST\ and SDSS, with the sources observed with \HST\ in dark blue, and those for which we are lacking \HST\ data shown in orange. In light grey, we show the bins from which \fpsfhst\ is estimated where we do not have \HST\ data. We take a bin surrounding the \fpsfsdss\ point, of width 0.2. Using the median \fpsfhst\ values from the points within that bin, we determine the equivalent \fpsfhst. Thus there is one light grey bin for every source lacking \HST\ data. The large error bars on the estimated \fpsfhst\ points are due to the large scatter. The fact that every point is either on or below the line of equivalence demonstrates that SDSS overestimates \fpsf, and hence we need \HST\ data.}
    \label{fig:fpsf}
\end{figure}

The availability of \HST\ imaging for part of \agndiscs\ facilitates more accurate structural decomposition of these sources than was originally possible using SDSS images. The full details of AGN host structural decomposition of the \HST\ images will be presented in a separate work (M. Fahey et al, in preparation). Briefly, we used the two-dimensional parametric image fitting program \galfit\ \citep{peng2002, peng2010} to simultaneously model the unresolved nucleus and extended galaxy for each of the sources in \agndiscs\ that has \HST\ imagery. Each image was background-subtracted, and the sky model fixed to zero. We constructed an empirical PSF in each band using background-subtracted images of isolated stars drawn from every observation in the \HST\ snapshot programme described above.

We estimated initial guesses for fit parameters, using \textsc{Iraf} and \textsc{SAOImage DS9} \citep{joye2003} to measure central source positions and galaxy effective radii, as well as galaxy position angles and axis ratios. Each source was initially fit in an iterative `batch' mode, starting with a single \citet{sersic1968} profile for the galaxy model and a PSF for the AGN model. The host S\'ersic index is set to $n = 2.5$ and allowed to vary. This value was chosen so as to avoid favouring either an exponential disc ($n = 1$) or a deVaucouleurs bulge ($n = 4$). Where present, we also fit and subtract nearby bright stars and extended companion galaxies, and mask fainter compact sources from the fit. Subsequent batch-fitting iterations of each source involve additional galaxy components, including a compact S\'ersic component to model a potential pseudo- or classical bulge. 

Following the completion of batch fitting, we followed up each source to refine the fit. Where justified by inspection of fit residuals and reduced $\chi^2_\nu$, we refined the original fits and/or added additional components, including bars and spiral arms. In a few cases where the AGN emission saturated the detector in the long \HST\ exposures, we determined the AGN-to-galaxy luminosity ratio using fits to the short-exposure images, fixing this AGN magnitude and masking out the saturated pixels in subsequent fits to the galaxy in the long-exposure images. The overall goal of the fits to each source was to neither over- nor under-subtract the galaxy’s central region. In addition, great care was taken to ensure the chosen galaxy best fit contains physically realistic component parameters. 

The final photometric fits were used to determine the fraction of the total flux of the source coming from the AGN, \fpsf. This was done by assuming that the PSF component measured from the \HST\ images, \fpsfhst, is wholly due to the AGN point source in the centre of the system. For systems where \HST\ imagery is available, \fpsf\ is then calculated by dividing the fitted PSF by the sum of fluxes from all components. Throughout this paper, when referring to the galaxy flux, this is the the total flux multiplied by $\left(1-f_{\mathrm{psf}}\right)$.

As mentioned {above}, \HST\ images are not available for the entire \agndiscs\ sample, and thus those sources lacking \HST\ data require us to estimate their individual values of \fpsf. All sources in \agndiscs\ have an estimate of \fpsf\ from SDSS. We calculate this value, \fpsfsdss , for all sources in \agndiscs\ using the \texttt{psfMag} and \texttt{cModelMag} SDSS photometric values to determine the PSF and total source flux, respectively. As discussed in \citet{simmons2017}, \fpsfsdss\ is overestimated for these systems given their bright nuclear emission and the resolution of SDSS compared to the size of the galaxies. Given that \HST\ has a factor of $\sim 8$ better resolution, we expect the \HST-derived values of \fpsf\ to be far more accurate. Figure \ref{fig:fpsf} shows the values of \fpsf\ from both SDSS and \HST\ for each system with available \HST\ images. The \fpsfsdss\ is higher than \fpsfhst\ for \emph{every} system, confirming the predictions of \citet{simmons2017}. Additionally, the 43 systems in \agndiscs\ with measurements from both SDSS and \HST\ allow us to determine a relation between the lower-resolution and higher-resolution measures, which we apply to the remaining systems without \HST\ data. Specifically, we determine a running median of the ratio between \HST\ and SDSS PSF flux fractions, using a sampling width of 0.2 in SDSS PSF flux fraction. We extrapolate this median, assuming a linear increase, for the 6 data points outside the range of values observed in the subset of \agndiscs\ with both \HST\ and SDSS measurements. For each source lacking an \HST\ image, we assume the \HST\ \fpsf\ is equal to the SDSS PSF fraction times the median ratio, with an uncertainty determined by sampling the scatter in the distribution at that value. The estimated values of \fpsf\ and their uncertainties are shown in Figure \ref{fig:fpsf}.

\subsection{Bar presence}\label{sec:bar_presence}
{There are several methods used to classify bars, most commonly via ellipse fitting \citep[e.g][]{regan1997}, and visually \citep[e.g.][]{nair2010}. The method used is unlikely to affect final counts, as demonstrated in \citet{sheth2008}, who used both methods to classify their sample of over 2000 face-on, spiral galaxies into strongly barred, intermediate barred and non-barred. They found that the two methods agreed in 85 per cent of cases, and in a further 10 per cent of cases, only disagreed by one class. A further method of bar classification is using GZ2, which classifies a galaxy's bar status in the same style as the identification of discs (Section \ref{sec:inactive_sample}). Once a volunteer has established that the source is a disc that is not edge-on, they are asked ``Is there a sign of a bar feature through the centre of the galaxy?''. GZ2 has been shown in multiple studies \citep[e.g.][]{masters2011, simmons2014} to robustly identify bars. \citet{melvin2014} used Galaxy Zoo Hubble (GZH), which follows the same question tree as GZ2, to investigate bar fraction with redshift, and their results are in strong agreement with \citet{sheth2008}. This shows that the three methods -- ellipse fitting, visual, and Galaxy Zoo -- can all be used in conjunction with each other to obtain robust classifications of bar status. Many previous GZ2 bar studies focus on strong bars, and thus use a relatively high threshold for bar selection (e.g., $p_{\mathrm{bar}} \geq 0.5$). \citet{willett2013} show that the optimal GZ2 vote fraction for including both strong and weak bars in an analysis of population bar fractions is $p_{\mathrm{bar}} \geq 0.3$}.

For \agndiscs, visual identification of a bar was performed by a single expert classifier (ILG) using the \HST\ images for the 43 sources that have such data available. The same classifier then repeated this visual identification for the 23 sources for which we are lacking \HST\ data using SDSS images of the galaxies. Only two galaxies in \agndiscs\ had been classified in GZ2, thus we did not use GZ2 to identify bar presence. We note that due to the brightness of the AGN, we may have missed some smaller bars that would still be classed as galactic-scale, and acknowledge that this is an additional source of asymmetric uncertainty, and thus the true bar fraction for this sample may be higher than we show.

{The bar status of all the galaxies in \inacdiscs\ was determined using a GZ2 bar vote fraction threshold of $p_{\mathrm{bar}} \geq 0.3$. A number of these were visually checked by ILG to ensure consistency with \agndiscs . The results presented in Section \ref{sec:controlling_for_sfr_and_mass} do not depend strongly on the vote fraction threshold.}
%We use a combination of methods to classify each galaxy as either having a bar or not having a bar. For \agndiscs, visual identification of a bar was performed by a single expert classifier (ILG) using the \HST\ images for the 43 sources that have such data available. The same classifier then repeated this visual identification for the 22 sources for which we are lacking \HST\ data using SDSS images of the galaxies. Only two galaxies in \agndiscs\ had been classified in GZ2, thus we did not use GZ2 to identify bar presence. We note that due to the brightness of the AGN, we may have missed some smaller bars that would still be classed as galactic-scale, and acknowledge that this is an additional source of asymmetric uncertainty, and thus the true bar fraction for this sample may be higher than we show.

%We used GZ2 to classify the inactive galaxies' bar status in the same style as the identification of discs. Once a volunteer has established that the source is a disc that is not edge-on, they are asked "Is there a sign of a bar feature through the centre of the galaxy?". \citet{willett2013} show that the optimal vote fraction for classifying a galaxy as barred, including both weak and strong bars, is $p_{\mathrm{bar}} \geq 0.3$, and we use this same threshold.

\subsection{Bulge Classification}\label{sec:bulge_presence}
We classify the galaxies in \inacdiscs\ into those containing a bulge at the centre of their disc, and those that have a bulge prominence no greater than that in \agndiscs, following the method outlined in \citet[Equation 3]{masters2019} to determine the bulge prominence, $B_{\mathrm{avg}}$ using GZ2. After deciding whether a disc galaxy has a bar, volunteers are asked `How prominent is the central bulge, compared with the rest of the galaxy?' and presented with four options: `No bulge', `Just noticeable', `Obvious', and `Dominant'. 

\begin{equation}
   B_{\mathrm{avg}} = 0.2p_{\mathrm{just\ noticeable}} + 0.8p_{\mathrm{obvious}} + 1.0p_{\mathrm{dominant}}
\end{equation}

By visually inspecting whether a subsample of galaxies are visually bulgeless, we determine what value of $B_{\mathrm{avg}}$ we require so that the bulge prominence parameter agrees with visual observations. A useful condition for a disc galaxy that is not edge-on to be classified as having a bulge prominence in line with \agndiscs\ is $B_{\mathrm{avg}} \leq 0.3$.

\section{Stellar properties of the samples}\label{sec:stellar_props}

Given that we need to control for SFR and stellar mass, \mstar, we first need to measure these parameters, and we describe this process below.  Figure \ref{fig:sfr_m*_graph} shows the SFR-\mstar\ distribution of the parent inactive sample, \inacdiscs\ (dark blue contours), and the complete disc-dominated, AGN host sample, \agndiscs\ (red crosses). The two samples, whilst they have significant overlap in their distributions, occupy very different parameter spaces. The process for obtaining \mstar\ is described in the Section \ref{sec:mstar}, and the process for obtaining SFR is described in Section \ref{sec:sfr}.

\subsection{Stellar Mass}\label{sec:mstar}

For the \inacdiscs\ sample, we use the median stellar mass value reported in the MPA-JHU catalogue \citep{kauffmann2003a,salim2007,brinchmann2004} for each individual galaxy. {This is possible since there are no bright AGN in the galaxies in \inacdiscs, so there is no need to account for the flux coming from the AGN contaminating the galaxy flux.} The minimum \mstar\ is $\log(M_{\ast}) = 7.20$ and the maximum SFR is $\log(M_{\ast}) = 12.06$. The median is $\log(M_{\ast}) = 9.80$.

{It is important that stellar mass is calculated in as similar way as possible for both samples. \citet{kauffmann2003a} used SDSS-derived spectral indices to determine stellar masses, correcting for a number of potential biases, including for the size and partial galaxy coverage of the spectral fibre aperture. They also found a tight relation between galaxy colour and mass-to-light ratio. The colour-based $M/L$ determination directly uses the integrated light of the whole galaxy. In addition to being generally useful for galaxies where no spectrum is available, this method is likely to be more robust to contamination from luminous AGN than the method based on fibre spectra.} %i don't like this. This is not a proper citation for the determination of M*
% Note salim2007 only do SFRs based on UV photometry, not stellar masses, so I've moved that ref up to the overall MPA-JHU ref.

We estimate \mstar\ for the \agndiscs\ sample using the colour-dependent mass-to-light ratio determinations of \citet[Figure 5]{baldry2006}. This method requires $u-r$ colours for the host galaxies, disentangled from the bright AGN emission. We assume that our measured \fpsf\ values (Section \ref{sec:HST_data}) are a better measure of AGN and host galaxy flux ratios than the SDSS \texttt{psfMag} in every band, and thus apply the factor of $(1 - f_{\mathrm{psf}})$ to the $u$ and $r$ band \texttt{cModelMag} to determine galaxy $u$ and $r$ magnitudes. The minimum \mstar\ is $\log(M_{\ast}) = 9.93$ and the maximum \mstar\ is $\log(M_{\ast}) = 11.19$. The median is $\log(M_{\ast}) = 10.71$.

From Figure \ref{fig:fpsf}, we can see that had we used exclusively \fpsf\ from SDSS, the values for \mstar\ would tend to be underestimated, since the fraction of the total flux assigned to the AGN would be greater than the true value, {leading to a lower flux being assigned to the galaxy. Following the equations in \citet{baldry2006}, this} would lead to a lower \mstar. Our improved PSF subtraction allows us to determine stellar masses for the AGN sample that more closely match the masses determined for the inactive sample. In Section \ref{sec:sfr_only} we also match the stellar mass distributions between \agndiscs\ and \inacdiscs .

\subsection{Star Formation Rate}\label{sec:sfr}
{As with \mstar\, it is important that the methods for calculating SFR in \agndiscs\ and \inacdiscs\ are as similar as possible, whilst acknowledging that only one sample has a source of flux of contamination in the form of an AGN.}

We use the formula outlined in \citet{kennicutt1994}, succinctly expressed in solar units in \citet[Equation 14]{pflamm-altenburg2007} to determine the SFR of individual galaxies in \agndiscs, where $L_\mathrm{H\alpha}$ is the H$\upalpha$ luminosity.

\begin{equation}
    \frac{\mathrm{SFR}}{\mathrm{M}_{\odot}\,\mathrm{yr}^{-1}} = \frac{L_\mathrm{H\alpha}}{1.26\times 10^{41}{\,\mathrm{erg}\,\mathrm{s}^{-1}}}
    \label{eq:Ha_SFR}
\end{equation}

However, this only gives the SFR within the region observed with Lick (see Figure \ref{fig:sdss_obs_region}), SFR$_{\mathrm{obs}}$, and requires extrapolation to the rest of the galaxy, SFR$_{\mathrm{gal}}$. We do this via simplification of the method outlined in \cite{brinchmann2004}, which assumes that SFR directly correlates with the luminosity in the $i$-band. We determine the $i$-band luminosity in the observed region, $L_{i,\mathrm{obs}}$, by convolving the spectrum with the $i$-band filter transmission curve \citep{rodrigo2012, rodrigo2020}. We use the SDSS \texttt{cModelMag} from the MPA-JHU catalogue to calculate the $i$-band luminosity of the galaxy, $L_{i,\mathrm{gal}}$ (via use of \fpsf), and scale up the SFR accordingly via:

\begin{equation}
    \mathrm{SFR}_{\mathrm{gal}} = \frac{L_{i,\mathrm{gal}}}{L_{i,\mathrm{obs}}}\mathrm{SFR}_{\mathrm{obs}}
    \label{eq:sfr_scale}
\end{equation}

Using SDSS flux measurements from MPA-JHU, the inactive sample is consistent with a single value of $0.3\pm0.1$ for the Balmer decrement, assuming a gas temperature of $T = 10^{4}\,\mathrm{K}$, an electron density of $n_e = 10^{2}\,\mathrm{cm}^{-2}$, Case B recombination \citep{osterbrock1989}, and a reddening curve defined in \citet{calzetti2000}. We assume that this also applies to the star-forming regions of the AGN-host galaxies, and thus apply this Balmer decrement as shown in \citet{dominguez2013}.

There are {22} sources in \agndiscs\ for which we were unable to obtain values of H$\upalpha$ flux in the galaxy, and can only constrain the upper limit. This is due to no discernible signal, even after carefully removing the AGN contamination from the galaxy using the wings of the PSF, as described in Section \ref{sec:spectral_fitting}. Thus, for galaxies that have an upper limit to their H$\upalpha$ flux, they only have an upper limit for their SFR.

Since the sources in \inacdiscs\ do not host a bright AGN contaminating the emission from the galaxy, we can directly use the values in MPA-JHU for {total} SFR {(as opposed to the SFR exclusively in the central fibre)}, given as \texttt{MEDIAN\_SFR}{, which also uses the method outlined in \citet{brinchmann2004}}. The minimum SFR is $\log(\mathrm{SFR}) = -2.40$ and the maximum SFR is $\log(\mathrm{SFR}) = 1.93$. The median is $\log(\mathrm{SFR}) = 0.026$.

\subsubsection{Dealing with upper limits}\label{sec:upper_limits}
We identify whether the {22} galaxies with no detected H$\upalpha$ emission are consistent within our S/N limits with being drawn from the subsample of 34 galaxies in \agndiscs\ with H$\upalpha$ detections. We use a bootstrapping method to randomly sample from within the upper limits of the non-detected SFRs. Specifically, we assume the true values of SFR are uniformly distributed between the upper limit calculated, and a lower end of $\log(\textrm{SFR}) = -1.5$, where $-1.5$ was chosen as a small, non-zero number approximately equal to the lower end of SFRs in \inacdiscs. A uniform distribution is a conservative estimate, since there is no reason to assume that the true value of the SFR is closer to the upper limit than to anywhere else in the range -- we have no prior information about the distribution of SFRs. We also select a random sample from the sources with H$\upalpha$ detections, where the SFR was randomly drawn from a normal distribution with a mean of $\log$(SFR) and a standard deviation of the error in $\log$(SFR). We re-sampled from upper limit SFRs and values of SFR using this method 100,000 times, with replacement. For each sampling, we used a KS test \citep{kolmogorov1933} to identify the probability that the two samples were drawn from the same distribution. If the SFRs of the limited subsample are statistically indistinguishable from those in the measured subsample, we would expect the KS values to follow a Normal distribution. For example, we would expect approximately 95 per cent of tests to have $p > 0.05$.

Instead, the distribution of KS values from the bootstrapping is highly skewed toward more statistically significant differences. Only 0.077 per cent of the selections and comparisons had $p > 0.05$. In other words, a $>2\upsigma$ confidence that the two samples were statistically indistinguishable only occurred 77 times out of 100,000. If the subsample with limits was indistinguishable from that without, we would expect this to occur approximately 95,000 times. Therefore the sources with SFR limits do have significantly lower SFRs than the rest of the sample, but our inability to otherwise constrain them inhibits a clean comparison with the inactive sample. Thus, for comparisons using a tightly controlled sample, we remove the sources which have only upper limits on their SFR, instead of H$\upalpha$ detections.

This gives us an AGN host galaxy sample used for comparison, which we call \agndiscfin, of median redshift 0.13, containing 34 galaxies, 20 of which host a large-scale galactic bar. The fraction of this sample hosting a bar is $f_{\mathrm{bar}, \mathrm{AGN}} = 0.59^{+0.08}_{-0.09}$, where uncertainties enclose the 68 per cent confidence limits of the binomial fraction error \citep{cameron2011}. The minimum SFR is $\log(\mathrm{SFR}) = -1.16$ and the maximum SFR is $\log(\mathrm{SFR}) = 2.08$. The median is $\log(\mathrm{SFR}) = 0.56$.

With both \mstar and SFR derived from AGN-subtracted galaxy fluxes, we can examine further the star-forming properties of the sample below.

\section{Star Formation in Merger-Free AGN hosts}\label{sec:sfr_only}
\begin{figure}
    \centering
    \includegraphics[width=\columnwidth]{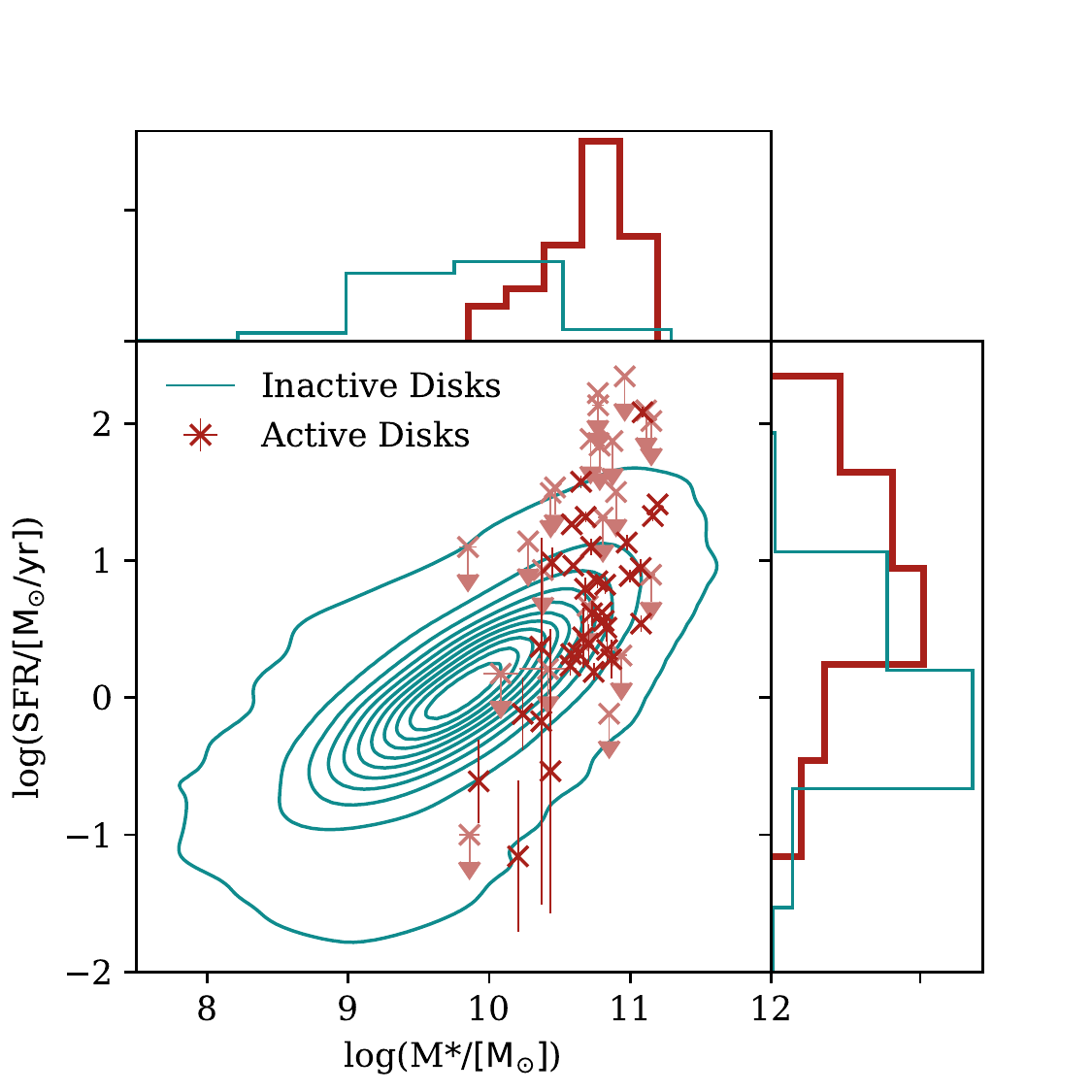}
    \caption{SFR against \mstar, for both the active sample, \agndiscs\ (red crosses) and the inactive disc-dominated sample, \inacdiscs\ (dark blue contours). Upper limits for SFR in the active sample are shown as arrows and in a slighter paler red than those with values. Normalised histograms are shown on the top and right axes, with the thick red line corresponding to \agndiscs\ and the thin blue line to \inacdiscs.}
    \label{fig:sfr_m*_graph}
\end{figure}

In order to examine SFRs in both the AGN host and inactive galaxy samples, we must first control for differences in stellar mass. Figure \ref{fig:sfr_m*_graph} shows that whilst there is considerable overlap in the two samples in their stellar mass distributions, the distributions remain noticeably different -- for active galaxies the distribution is narrower than for inactive galaxies, with the average active galaxy's \mstar\ lying above the median \mstar\ of inactive galaxies. This pattern remains upon the removal of the galaxies with only upper limits on their star formation rate.

The difference in \mstar\ between the two samples is most likely due to selection effects rather than an intrinsic difference. \agndiscs\ is selected as a sample to host only the most luminous AGN. If we assume that the sample is not as a whole exceeding the Eddington limit, this means that there is a lower limit on black hole mass, $M_{\mathrm{BH}}$. It is broadly understood that there is some form of co-evolution between galaxies and SMBHs \citep[e.g.][]{kormendy2013}, even if we continue as a field to debate the details. Thus a lower limit on $M_{\mathrm{BH}}$ implies a lower limit on \mstar, and the sample is therefore self-limiting regarding \mstar\ \citep[for a deeper exploration of this selection bias, see][]{aird2012}.

The other way that \agndiscs\ self-limits in \mstar\ is that the sample is selected to consist of strongly disc-dominated galaxies. The galaxies were identified using SDSS, where the PSF width may be a substantial fraction of a galaxy's extent. If a low-mass disc-dominated galaxy hosted a very luminous AGN, the AGN would outshine the galaxy and the disc would be difficult or impossible to identify in SDSS imagery at the redshifts of this sample. Such a galaxy would not be included in \agndiscs. Therefore there is a lower limit on disc radius, which implies a lower limit on \mstar.

These two selection effects mean we have very few AGN hosted in galaxies with $M_{\ast}\,<10^{10}\, \mathrm{M}_{\odot}$ in our sample, and hence we must select galaxies from \inacdiscs\ which have the same \mstar\ distribution before comparing SFRs between the samples.

\begin{figure*}
    \centering
    \begin{subfigure}{0.47\textwidth}
        \centering
        \includegraphics[width=\columnwidth]{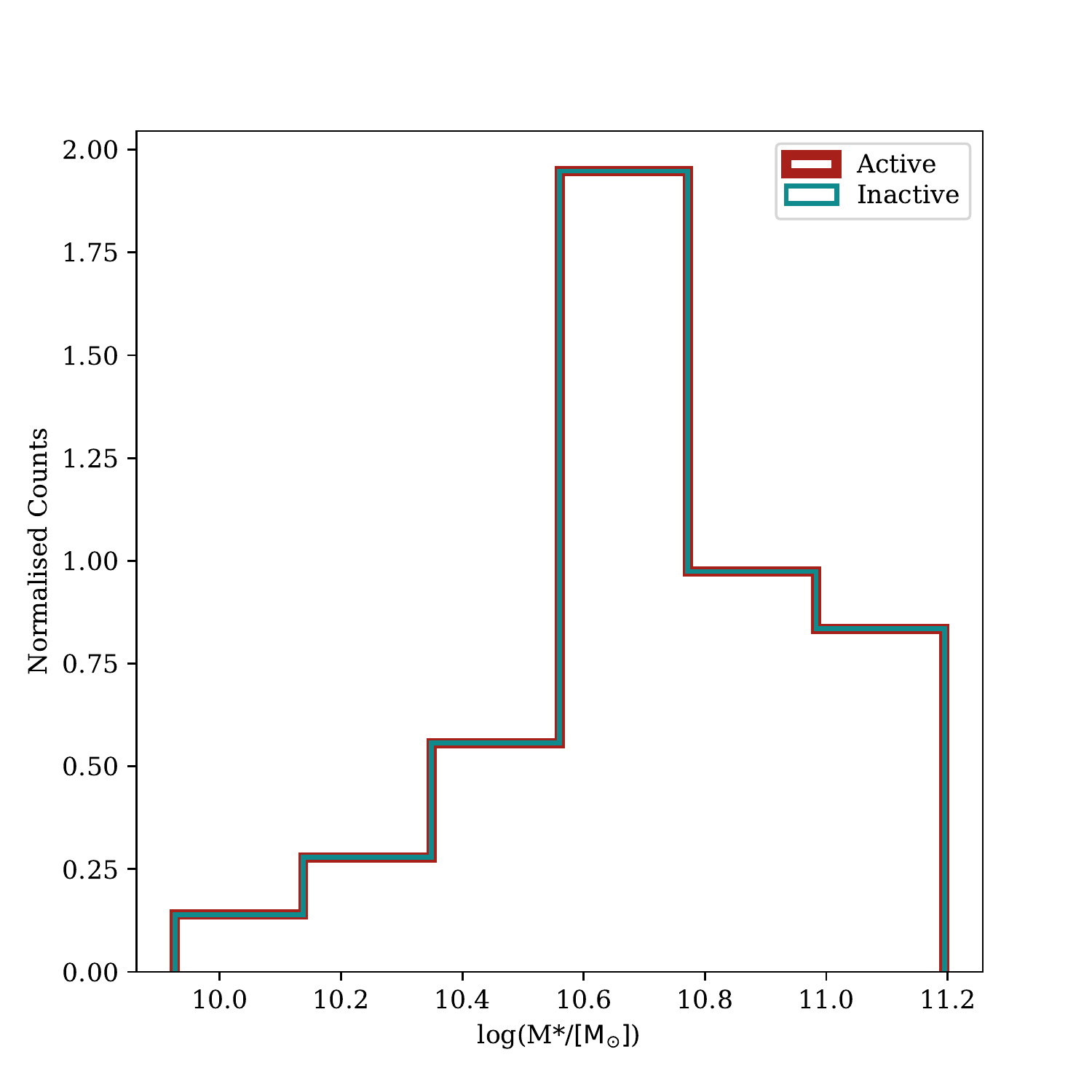}
        \caption{}
        \label{fig:disc_mass_hist_m*_cont}
    \end{subfigure}
    \begin{subfigure}{0.47\textwidth}
        \centering
        \includegraphics[width=\columnwidth]{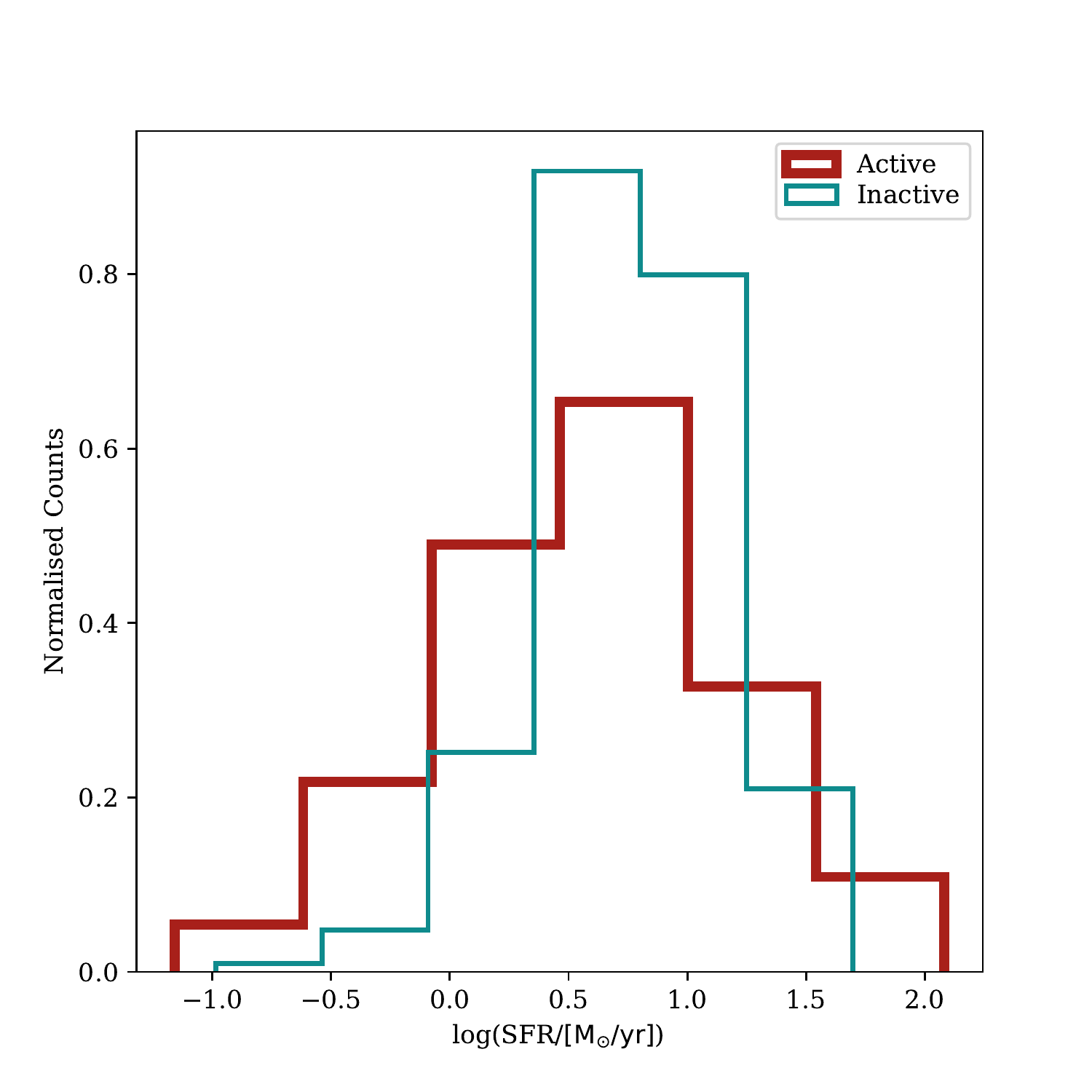}
        \caption{}
        \label{fig:disc_sfr_hist_m*_cont}
    \end{subfigure}
    \caption{Distribution of \mstar\ (left panel) and SFR (right panel) after controlling for \mstar, with AGN host galaxies shown in thick red lines and inactive galaxies shown in thin blue lines. The \mstar\ distribution demonstrates that we have successfully controlled for \mstar, and has a $p$-value from a KS test of 1.000, showing the samples are consistent with being drawn from the same parent sample. The SFR histogram also shows the similarity between the two samples after controlling for \mstar, and with a $p$-value of 0.368, the SFR's are consistent with being drawn from the same parent sample.}
    \label{fig:disc_hist_m*_cont}
\end{figure*}

We control for \mstar\ by weighting the inactive sample in six bins of equal width. This \mstar-matched subset of inactive disc galaxies is hereafter called the \inacdiscmatch\ sample, and its \mstar\ distribution is shown in Figure \ref{fig:disc_mass_hist_m*_cont}, for comparison with \agndiscfin.  After performing a KS test on \agndiscfin's and \inacdiscs's \mstar\ to confirm their similarity, we obtain a $p$-value of $p_{mass} = 1.000$, which demonstrates that \agndiscfin\ and \inacdiscmatch\ are consistent with being drawn from the same parent sample.\footnote{All reported $p$-values for KS tests between weighted distributions are estimated using sample weights instead of raw object counts.} The distribution of SFRs for the \mstar-matched \agndiscfin\ and \inacdiscmatch\ samples are shown in Figure \ref{fig:disc_sfr_hist_m*_cont}. The slight visual differences between the distributions do not appear to be statistically significant (KS $p_{\mathrm{SFR}} = 0.368$, a significance of $0.9 \upsigma$). Thus we cannot rule out the null hypothesis that the SFRs of these disc-dominated galaxies hosting luminous Type-1 AGN are drawn from the same parent population as a sample of disc-dominated galaxies \emph{not} hosting AGN. Our qualitative results do not change if we instead draw \mstar-matched sub-samples instead of weighting the respective distributions.

While a lack of statistically significant differences in the SFRs of these subsamples may be due to our relatively small sample sizes, we might expect such a result even with a larger sample, due to the complex physical processes at play. For example, AGN outflows may both quench and enhance star formation in a host galaxy (see \citealt{harrison2017} for a review). A high fraction of our AGN host sample shows signs of outflows \citep{smethurst2019, smethurst2021}, and these galaxies do not congregate in a specific region of SFR--\mstar\ space (Figure \ref{fig:sfr_m*_graph}), consistent with expectations. Differing timescales also complicate interpretation of our results: the duration over which an AGN is active in a galaxy may be considerably shorter than the effects of AGN-driven quenching \citep{schawinksi2015}, which would further dilute differences between SFR in the AGN host and inactive disc galaxy population. Better constraints on population differences between disc-dominated AGN host and inactive galaxies will require a larger sample, ideally with spatially-resolved spectral information to more robustly trace the effect of AGN feedback.

\section{Bar Fractions of AGN host vs Inactive discs} \label{sec:controlling_for_sfr_and_mass}
In order to isolate the possible effect of large-scale, galactic bars, we first need to ensure that all other variables which are known to correlate with bar fraction are negated via careful weighting in \mstar\ and SFR to obtain a comparison sample. We use the star-forming sequence shown in Figure \ref{fig:sfr_m*_graph} to ensure that both the active and inactive samples are consistent with each other in their \mstar\ and SFR, an additional control compared to Section \ref{sec:sfr_only}, where we only control for \mstar. As with \mstar, there is significant overlap in SFR between the two samples. Whilst the SFR for active galaxies seems to cover approximately the same range as that for inactive galaxies, when we only control for \mstar\ the samples still differ enough in SFR that we need to control for SFR in order to analyse the bar fraction. The medians of the two samples are $\mathrm{SFR}_{\mathrm{AGN}} = 0.59$ and $\mathrm{SFR}_{\mathrm{inactive}} = 0.72$, and the ranges are $ -1.16 \leq \mathrm{SFR}_{\mathrm{AGN}} \leq 2.18$ and $ -0.62 \leq \mathrm{SFR}_{\mathrm{inactive}} \leq 1.69$. Given that the two samples have different distributions, it is vital that we control for SFR as well as \mstar, in order to truly isolate the effect of the bar.

We divide the \mstar\ and SFR each into six bins, and assign weights to each galaxy in \inacdiscs, such that the weighted sample (which we hereafter call \inacdiscmatch) has \mstar\ and SFR distributions matching those of \agndiscfin. This gives a weighted bar fraction for \inacdiscmatch\ of $f_{\mathrm{bar, Inac}} = 0.44^{+0.08}_{-0.09}$, where uncertainties arise from the binomial fraction error \citep{cameron2011}.

\begin{figure*}
    \centering
    \begin{subfigure}{0.47\textwidth}
        \centering
        \includegraphics[width=\columnwidth]{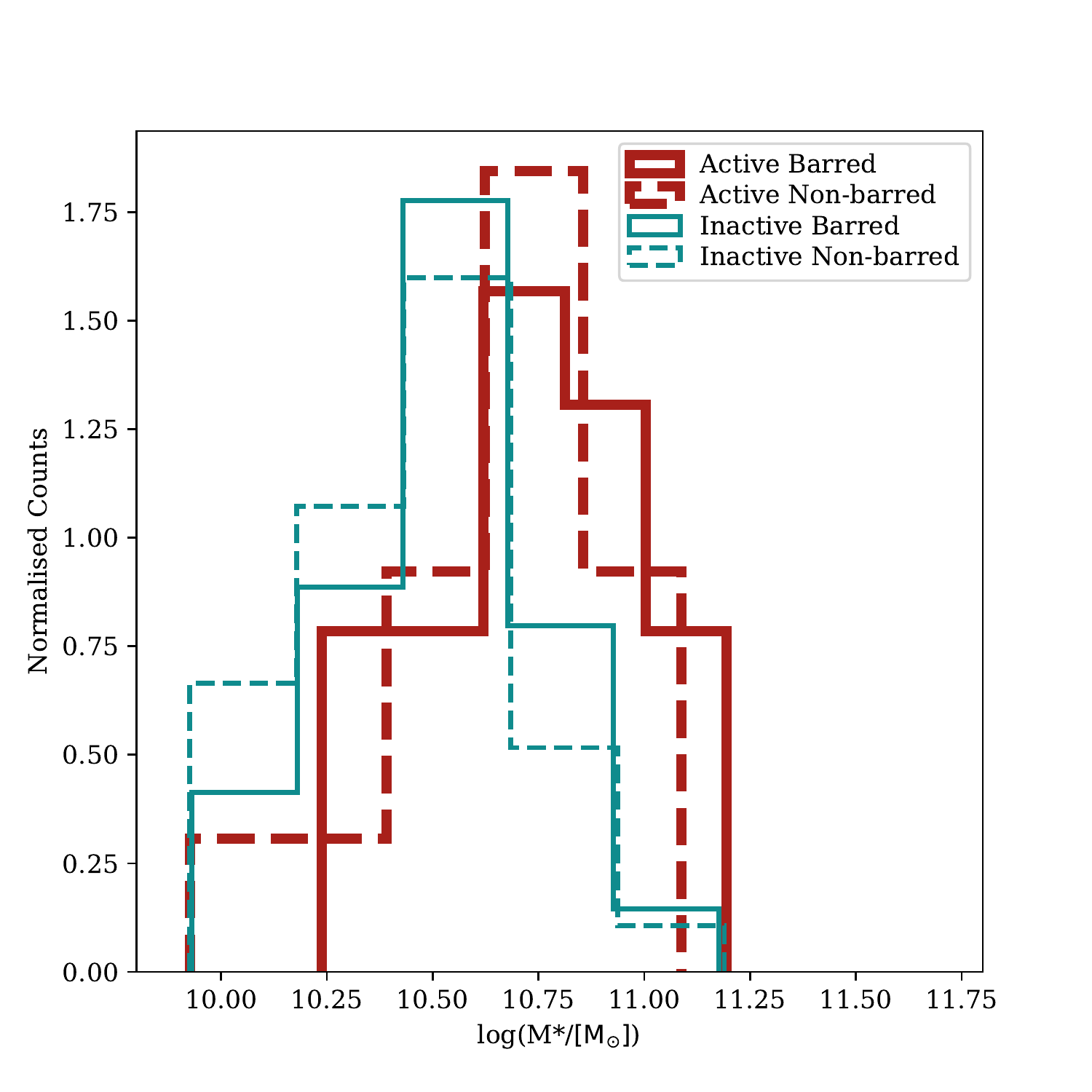}
        \caption{}
        \label{fig:mass_hist}
    \end{subfigure}
    \begin{subfigure}{0.47\textwidth}
        \centering
        \includegraphics[width=\columnwidth]{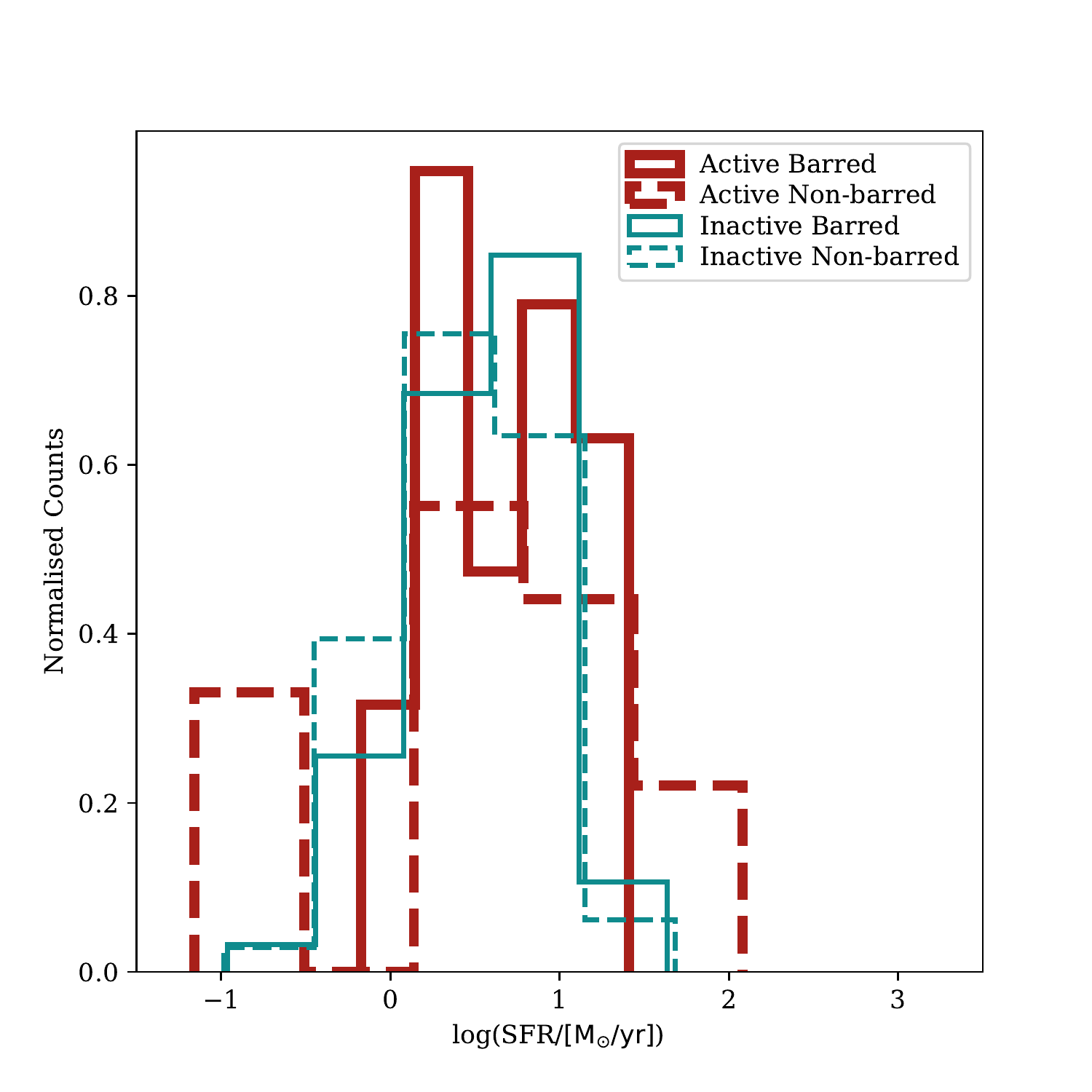}
        \caption{}
        \label{fig:sfr_hist}
    \end{subfigure}
    \caption{Distributions of \mstar\ (left panel) and SFR (right panel), after controlling for both of these parameters, split by active (red, thick line) and inactive (thin blue line) galaxies, and by barred (solid line) and non-barred (dashed) lines. The results of KS tests between each pair of samples are shown in Table \ref{tab:ks_test}, but all the samples are consistent with being drawn from the same parent sample in both \mstar\ and SFR.}
    \label{fig:hist}
\end{figure*}

We show the distributions of the control samples, split by active/inactive and by barred/non-barred, with \mstar\ in Figure \ref{fig:mass_hist}, and SFR in Figure \ref{fig:sfr_hist}. As expected, the distributions cover a much more similar range than in Figure \ref{fig:sfr_m*_graph}. We confirm via KS tests on \agndiscfin\ and \inacdiscmatch\ that their \mstar\ and SFR distributions are consistent with being drawn from the same parent sample.

We also use KS tests to compare both the SFR and the \mstar\ for different subsets of the comparison samples -- active galaxies, inactive galaxies, barred galaxies and non-barred galaxies. Table \ref{tab:ks_test} shows the $p$-values that result from the comparison samples in the first column. Values for the inactive subsamples are the weighted numbers.

\begin{table}
    \caption{KS test $p$-values from the comparisons described in Sections \ref{sec:sfr_only} and \ref{sec:controlling_for_sfr_and_mass}. These values are all indicative of statistically indistinguishable samples. Bold values indicate where we intentionally control for the samples to be statistically indistinguishable.}
    \label{tab:ks_test}
    \begin{tabular}{lcc}
        \hline
        Samples being compared & $p_{\mathrm{mass}}$ & $p_{\mathrm{SFR}}$\\
        \hline\hline
        Controlling only for stellar mass \\
        \hline
        \agndiscfin\ ($34$), \inacdiscmatch\ ($34$) & $\mathbf{1.000}$ & $0.368$\\
        AGN Bar ($20$), AGN Non-bar ($14$) & $0.814$ & $0.648$\\
        Inac Bar ($15$), Inac Non-bar ($19$) & $1.000$ & $1.000$\\
        AGN Bar ($20$), Inac Bar ($15$) & $1.000$ & $0.554$\\
        AGN Non-bar ($14$), Inac Non-bar ($19$) & $1.000$ & $0.710$\\
        \hline\hline
        Controlling for stellar mass and SFR \\
        \hline
        \agndiscfin\ ($34$), \inacdiscmatch\ ($34$) & $\mathbf{1.000}$ & $\mathbf{1.000}$\\
        AGN Bar ($20$), AGN Non-bar ($14$) & $0.814$ & $0.648$\\
        Inac Bar ($15$), Inac Non-bar ($19$) & $1.000$ & $0.977$\\
        AGN Bar ($20$), Inac Bar ($15$) & $1.000$ & $0.984$\\
        AGN Non-bar ($14$), Inac Non-bar ($19$) & $0.999$ & $0.955$\\
        \hline
    \end{tabular}
\end{table}

Looking at the bar fractions ($f_{\mathrm{bar,AGN}} = 0.59^{+0.08}_{-0.09}$ for \agndiscfin\ and $f_{\mathrm{bar,Inac}} = 0.44^{+0.08}_{-0.09}$ for \inacdiscmatch), we can see that after controlling for the SFR and \mstar, the sources in \inacdiscmatch\ are marginally less likely ($\sim 1.7 \upsigma$) to host a bar than the sources in \agndiscfin, in agreement with studies such as {\citet{alonso2013} and \citet{galloway2015}}. However it is worth noting that the samples used by \citet{galloway2015} contain $\sim10^5$ galaxies and this work contains $\sim10^2$ galaxies, yet both studies obtain a similar level of significance in their results. This could potentially be due to the fact that we are looking at galaxies with little--to--no bulge component, so any bars we have are likely to be younger than in \citet{galloway2015} where they make no distinction on bulge component, and thus we do not require such a large sample to obtain a similarly significant result. Our sample also considers only the highest luminosity AGN, whereas again, \citet{galloway2015} impose no such limit on their sample.

We can use the $p$-values from the KS tests shown in the second section of Table \ref{tab:ks_test} to rule out the null hypothesis that two samples are drawn from the same parent distribution. The first line, comparing \agndiscfin\ to \inacdiscmatch\ before controlling for SFR shows that overall the comparison samples are consistent with being drawn from the same parent sample. This is a simple check to confirm we have controlled for the various parameters correctly. From here, we divide each sample into barred and non-barred subsamples in order to draw comparisons.

For any \mstar- and SFR-matched sub-samples we examine, we cannot rule out the null hypothesis that the two samples are drawn from the same parent distribution. Several potential insights emerge from this overall result. Firstly, within our samples, a bar does not necessarily have to be present to form an AGN, but if there is a bar there, then it has no unique further effect on the SFR and \mstar. Secondly, the bar has no effect on SFR or \mstar\ in this SFR--\mstar\ regime. Lastly, barred AGN host galaxies are not a special subset of inactive barred galaxies, and this is mirrored by the comparison of active non-barred galaxies versus inactive non-barred galaxies, which also has $p$-values of SFR and \mstar\ close to 1, i.e., far short of any reasonable threshold for statistically significant differences. {This is much the same as results from works in the last few decades \citep[e.g.][]{ho1997, mulchaey1997, knapen2000, martini2003}. We would note that our results do not qualitatively change if instead we only consider strong bars in both samples (i.e., excluding weak bars in the AGN host sample and using a threshold of $p_{\mathrm{bar}}\geq0.5$ for the inactive sample to select strong bars, as described in Section \ref{sec:bar_presence}).}

It is worth noting that whilst these results indicate solutions, our sample of AGN hosts being used to quantitatively compare is simply too small to draw conclusions with much statistical power. This is because these are the very brightest AGN in the most unambiguously disc-dominated host galaxies, rather than a sample taken over the entire AGN population in all merger-free hosts. A significant portion of our sample has only upper limits on their SFR, further constraining the sample size. Our analysis of those limits (Section \ref{sec:upper_limits}) hints that higher signal--to--noise spectra permitting robust measurements of this subsample could provide further insight into our current results. Integral field spectroscopy for a large fraction of our sample would enable us to probe these galaxies in further detail, as would increasing the sample size by adding Vera Rubin Observatory’s LSST survey \citep{ivezic2019}, or getting more galaxies with Euclid or Roman. Since we are looking at a rare phenomenon (luminous AGN), in a rare subset of galaxies (bulgeless or nearly so), it really is important that we have a large volume so as to control for confounding variables and achieve statistically robust sample numbers. It is also crucial to remember that not all AGN are this luminous, this is a particular subset of AGN, and it was collected in such a way so as to show the possibilities of extreme conditions, and further data on less luminous AGN is needed to draw conclusions over the entire population.

\section{Conclusions}\label{sec:conclusions}
We have used a sample of unambiguously disc-dominated galaxies hosting luminous, Type-1 AGN in order to isolate SMBH growth through merger-free processes. We obtained longslit Lick spectroscopic data of the sample, and \HST\ images of part of the sample. This allowed us to measure robust SFRs and stellar masses for 34 galaxies -- the rest of the sample has only upper limits on their SFR. We compared this sample to a sample of inactive, disc-dominated galaxies with morphological classifications from Galaxy Zoo 2, and SFRs and \mstar\ from MPA--JHU. We performed KS tests on subsets of these samples, and we here summarise our findings:
\begin{itemize}
    \item Galaxies hosting an AGN have a wider range of SFR than galaxies lacking an AGN, with the SFR peaking at a slightly higher value.
    \item After controlling for SFR and \mstar, bars are marginally more likely to reside in AGN host galaxies than galaxies not hosting AGN, ($f_{\mathrm{bar}} = 0.59^{+0.08}_{-0.09}$ for \agndiscfin\ and $f_{\mathrm{bar}} = 0.44^{+0.08}_{-0.09}$ for \inacdiscmatch) -- there is a $\sim 1.7 \upsigma$ difference.
    \item Despite the fact that bars are more likely to reside in massive galaxies, and AGN are more likely to reside in massive galaxies, having both a bar and an AGN is not associated with a further increase in a galaxy's stellar mass beyond only having one of either a bar or an AGN.
\end{itemize}

Further work is needed to obtain higher resolution spectra for those galaxies where the flux from the disc is so overpowered by the flux of the AGN that we can only obtain upper limits of their SFR. This will allow for better separation of the AGN and the galaxy, which will result in a higher signal--to--noise ratio, and allow us to constrain SFRs further.

Upcoming surveys such as LSST and Euclid will facilitate breakthroughs in the field due to their increased resolution and sky coverage, which will allow us to obtain larger samples of merger-free AGN host galaxies for improved statistical analysis. With today's facilities and scientific ability, it is interesting to see that despite probing the extremes of black hole growth in the merger-free regime, for those galaxies where we can obtain SFR, they do not appear to be outliers compared to galaxies not hosting AGN.

\section*{Acknowledgements}
We would like to thank the anonymous referee for their valuable comments and insight, which improved the quality of this paper.

ILG acknowledges support from an STFC PhD studentship [grant number ST/T506205/1] and from the Faculty of Science and Technology at Lancaster University. BDS and KW acknowledge support through a UK Research and Innovation Future Leaders Fellowship [grant number MR/T044136/1]. RJS acknowledges funding from Christ Church, University of Oxford. TG acknowledges funding from the University of Oxford Department of Physics and the Saven Scholarship. DOR acknowledges support from an STFC PhD studentship [grant number ST/T506205/1]. MRT acknowledges support from an STFC PhD studentship [grant number ST/V506795/1].

This research is in part based on observations made with the NASA/ESA Hubble Space Telescope, obtained at the Space Telescope Science Institute, which is operated by the Association of Universities for Research in Astronomy, Inc., under NASA contract NAS5-26555. These observations are associated with program HST-GO-14606.

Support for program HST-GO-14606 was provided by NASA through a grant from the Space Telescope Science Institute, which is operated by the Association of Universities for Research in Astronomy, Inc., under NASA contract NAS5-26555.

This research has made use of the Spanish Virtual Observatory (\url{https://svo.cab.inta-csic.es}) project funded by MCIN/AEI/10.13039/501100011033/ through grant PID2020-112949GB-I00 \citep{rodrigo2012, rodrigo2020}.

Funding for SDSS-III has been provided by the Alfred P. Sloan Foundation, the Participating Institutions, the National Science Foundation, and the U.S. Department of Energy Office of Science. The SDSS-III web site is \url{http://www.sdss3.org/}.

SDSS-III is managed by the Astrophysical Research Consortium for the Participating Institutions of the SDSS-III Collaboration including the University of Arizona, the Brazilian Participation Group, Brookhaven National Laboratory, Carnegie Mellon University, University of Florida, the French Participation Group, the German Participation Group, Harvard University, the Instituto de Astrofisica de Canarias, the Michigan State/Notre Dame/JINA Participation Group, Johns Hopkins University, Lawrence Berkeley National Laboratory, Max Planck Institute for Astrophysics, Max Planck Institute for Extraterrestrial Physics, New Mexico State University, New York University, Ohio State University, Pennsylvania State University, University of Portsmouth, Princeton University, the Spanish Participation Group, University of Tokyo, University of Utah, Vanderbilt University, University of Virginia, University of Washington, and Yale University.

The data in this paper are the result of the efforts of the Galaxy Zoo volunteers, without whom none of this work would be possible. Their efforts are individually acknowledged at \url{http://authors.galaxyzoo.org}.

\subsection*{Software}
This research has made use of \textsc{Topcat} \citep{taylor2005}, an interactive graphical tool for analysis and manipulation of tabular data.

This research has made extensive use of the following Python packages:
\begin{itemize}
\item \textsc{Astropy}, a community-developed core Python package for Astronomy \citep{robitaille2013, price-whelan2018}.
\item \textsc{Matplotlib}, a 2D graphics package for Python \citep{hunter2007}.
\item \textsc{Numpy} \citep{harris2020}, a package for scientific computing.
\item \textsc{Scipy} \citep{virtanen2020}, a package for fundamental algorithms in scientific computing.
\end{itemize}

This research has made use of \textsc{Iraf} \citep{tody1986, tody1993} and its packages for longslit data reduction.

This research has made use of \textsc{Galfit} \cite{peng2002, peng2010} for fitting photometric data.

%%%%%%%%%%%%%%%%%%%%%%%%%%%%%%%%%%%%%%%%%%%%%%%%%%
\section*{Data Availability}
The data for \agndiscs\ is available on request.
%The inclusion of a Data Availability Statement is a requirement for articles published in MNRAS. Data Availability Statements provide a standardised format for readers to understand the availability of data underlying the research results described in the article. The statement may refer to original data generated in the course of the study or to third-party data analysed in the article. The statement should describe and provide means of access, where possible, by linking to the data or providing the required accession numbers for the relevant databases or DOIs.



\bibliographystyle{mnras}
\bibliography{bibliography} % if your bibtex file is called example.bib

%\begin{thebibliography}{99}
%\end{thebibliography}

%%%%%%%%%%%%%%%%%%%%%%%%%%%%%%%%%%%%%%%%%%%%%%%%%%

%%%%%%%%%%%%%%%%% APPENDICES %%%%%%%%%%%%%%%%%%%%%

\appendix

\section{SDSS thumbnails}\label{app:sdss_imgs}
\begin{figure*}
    \centering
    \includegraphics[width=0.850\textwidth]{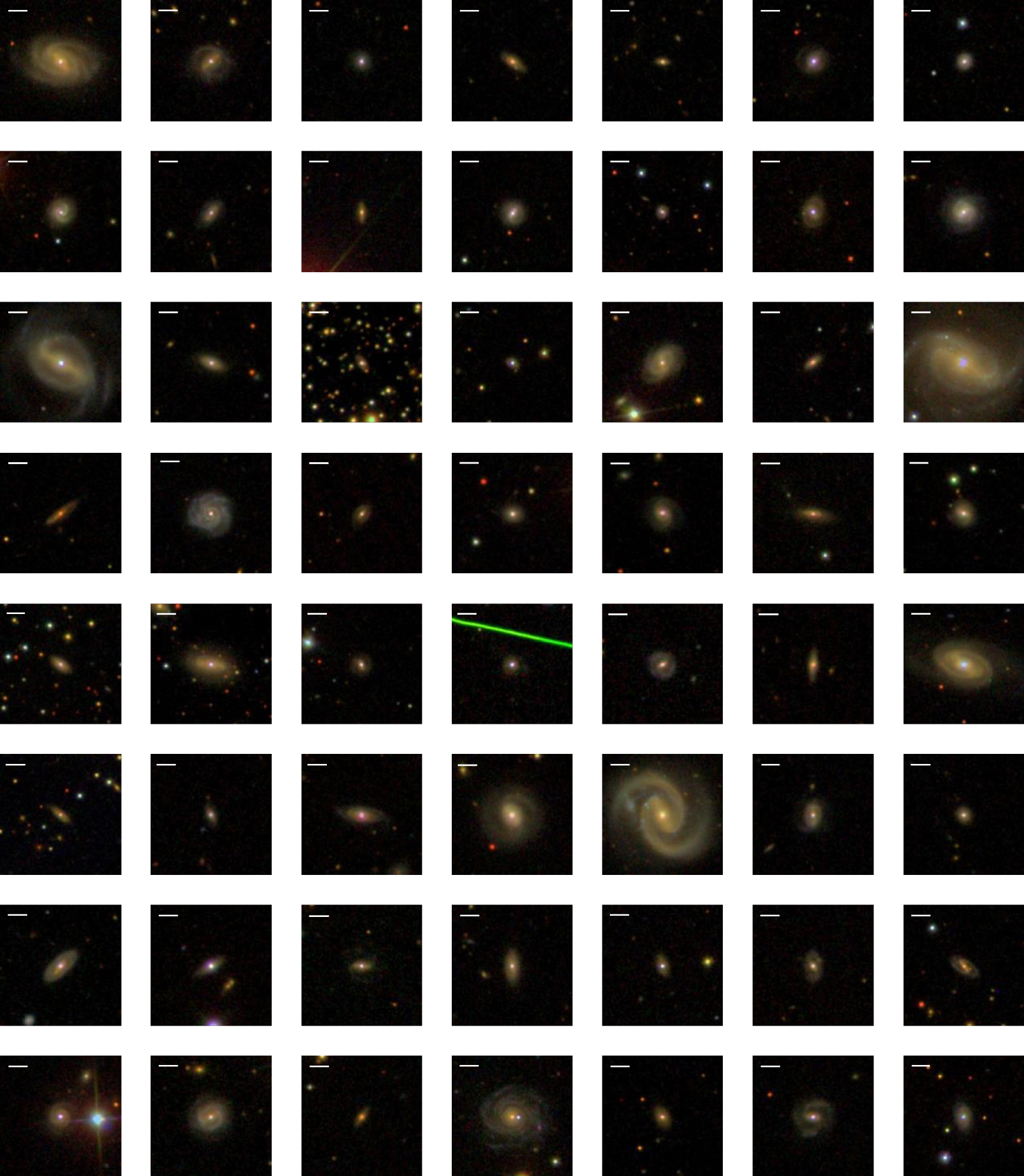}
    \caption{SDSS postage stamps of all galaxies in \agndiscs, including those that are constrained only by an upper limit in H$\upalpha$, and those with values. Images are taken from SDSS DR8 \citep{aihara2011}. The scale bar in each upper left corner represents 10 arcsec.The position angles of the galaxies do not correspond to those in Figure \ref{fig:all_hst}.}
    \label{fig:all_sdss}
\end{figure*}

Figure \ref{fig:all_sdss} shows the full \agndiscs\ sample imaged in SDSS, with the scale bar in each image representing 10 arcsec. The disc-dominated nature of the galaxies can be seen clearly, as well as a large-scale galactic bar in some images.

\section{\HST\ thumbnails}\label{app:hst_imgs}
\begin{figure*}
    \centering
    \includegraphics[width=0.850\textwidth]{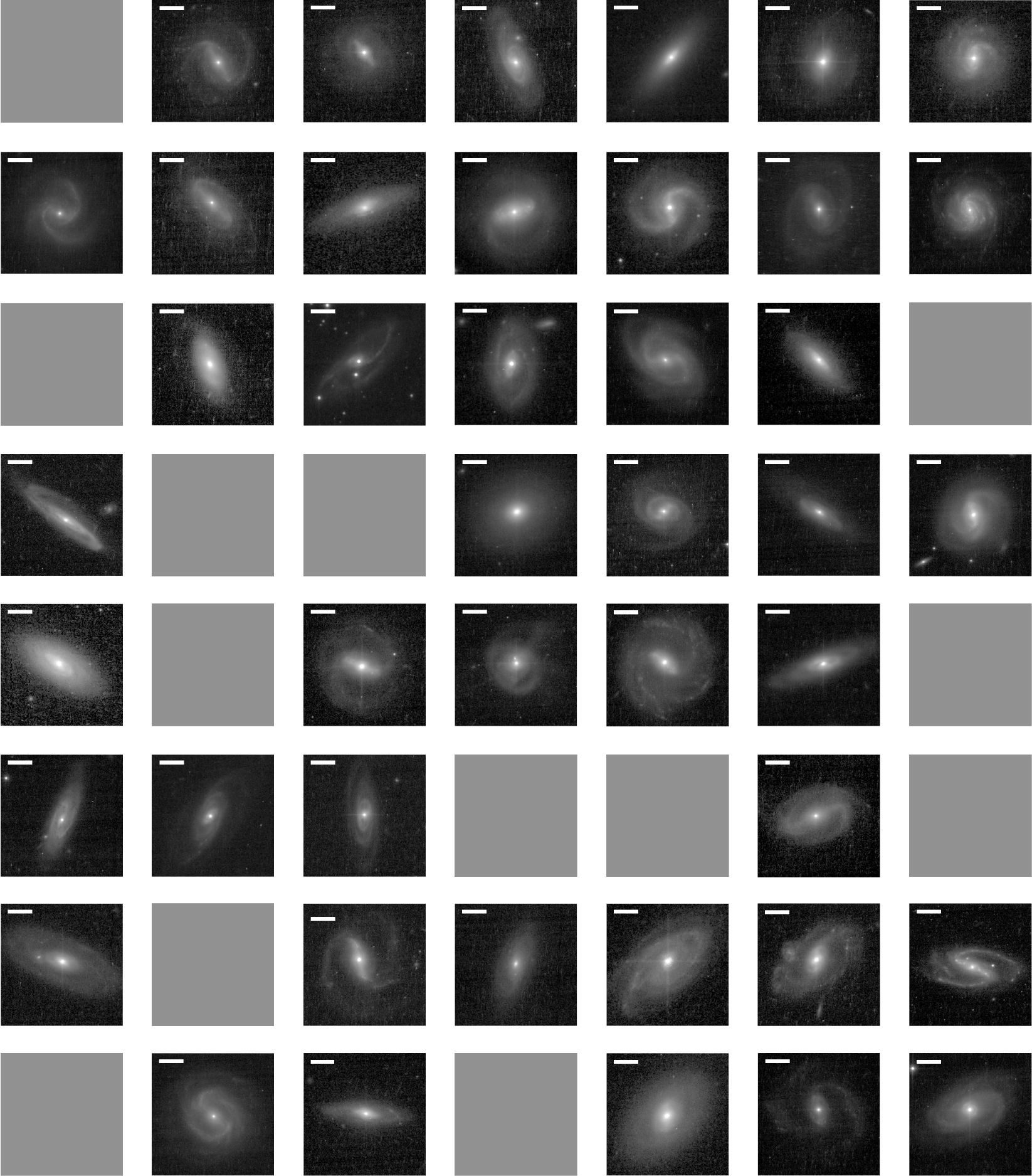}
    \caption{\HST\ postage stamps of the galaxies in \agndiscs\ that have been imaged in \HST. The galaxies' positions correspond to the galaxies in Figure \ref{fig:all_sdss}, and so the grey squares represent galaxies that have not yet been imaged with \HST. The white scale bar in each top left corner represents 5 arcsec. The position angles of the galaxies do not correspond to those in Figure \ref{fig:all_sdss}.}
    \label{fig:all_hst}
\end{figure*}

Figure \ref{fig:all_hst} shows the galaxies for which we have \HST\ data. Their position in the grid corresponds to their SDSS counterpart in Figure \ref{fig:all_sdss}, however their rotation does not. The scale bar in these images corresponds to 5 arcsec. The grey blank squares show galaxies for which we do not have \HST\ photometric data.

%%%%%%%%%%%%%%%%%%%%%%%%%%%%%%%%%%%%%%%%%%%%%%%%%%

% Don't change these lines
\bsp	% typesetting comment
\label{lastpage}
\end{document}